\documentclass[preprint,12pt]{elsarticle}
\usepackage{graphicx,epsfig,amssymb,amsmath,amsthm,amscd,amsfonts,bm,accents,hyperref} 
\usepackage[T1]{fontenc}    
\usepackage[utf8]{inputenc} 

\usepackage{float}
\usepackage{graphicx}
\usepackage{pifont}
\usepackage{stmaryrd}
\usepackage{wasysym}
\usepackage{marvosym}
\usepackage{bm}
\usepackage{appendix}
\usepackage{enumitem}
\usepackage[T1]{fontenc}
\usepackage[utf8]{inputenc}
\usepackage{yfonts}
\usepackage{textcomp}
\usepackage{lingmacros}
\usepackage[usenames]{color}

\usepackage[mathscr]{eucal}
\usepackage{comment}
\usepackage{multirow}
\usepackage{mathtools}
\usepackage{fixmath}
\usepackage{cuted}
\usepackage{url}
\biboptions{sort&compress}
\DeclareUnicodeCharacter{00FC}{\"u}

\usepackage{listings}
\usepackage{xcolor} 
\usepackage{hyperref}

\newcommand{\sa}[1]{\textcolor{blue}{#1}}
\newcommand{\hl}[1]{\textcolor{black}{#1}}

\usepackage[normalem]{ulem} 

\journal{Tribology International}

\begin{document}

\title{Plug Flow and Cavitation in Rough Lubricated Contacts: Molecular Dynamics of Single- vs. Two-Component Fluids}

\author[1]{Shubham Agarwal}

\author[1]{Martin H. Müser\corref{cor1}}
\ead{martin.mueser@mx.uni-saarland.de}
\affiliation[1]{
  organization={Materials~Science~and~Engineering},
  addressline={Campus~C6.3},
  city={Saarbr{\"u}cken},
  postcode={66123},
  country={Germany}
}


\cortext[cor1]{Corresponding author}



\begin{abstract}
We present non-equilibrium molecular dynamics simulations of lubricated sliding between rough, deformable surfaces under conditions representative of mixed lubrication.
One aim is to reduce the gap between highly idealized simulations of smooth interfaces and real, rough, load-bearing contacts.
Another aim is to determine whether favorable tribological properties of two-fluid lubrication reported for solvated hydrophilic-hydrophobic polymer-brush interfaces can also be realized in rough contacts without brushes.
To this end, we compare aqueous (water), hydrocarbon ($n$-dodecane), 
and immiscible two-fluid lubrication under identical geometric conditions \hl{representing randomly rough surfaces}.
For the single-component lubricants, the simulations reproduce established trends: 
Water shows stronger speed dependence but reduced load-bearing capacity than $n$-dodecane, despite their similar ambient viscosities.
%
Beyond this expected behavior, the simulations reveal that the combination of strong confinement and large height gradients can cause plug flow and \hl{thereby} cavitation after asperity collisions.
For a high-surface-tension liquid like water, cavitation leads to an abrupt shear-stress release on scales much exceeding the size of the cavity.
\hl{Since plug motion only weakens with single-component lubricants at low speeds, the two-fluid lubricant can only leverage its potential advantages at low sliding velocities.}
%
%
It is also the only system in which folding lips form, occasionally developing into transient, \hl{detached clusters} at high speeds.
\end{abstract}


\maketitle
\section{Introduction}

Friction and wear in metallic contacts are controlled by multiscale roughness: the applied load is carried by a sparse population of asperity junctions that experience high local pressures, strong adhesion, and pronounced plasticity~\cite{Bowden2001book,Persson2006SSR}. Lubricants mitigate these effects by redistributing load, separating the solids, and reducing interfacial shear forces.
Yet, friction and wear remain.
They depend sensitively on operating conditions and the properties of the lubricants—especially in mixed lubrication, where load is carried by both a sheared lubricant and solid-solid junctions.

Modern contact-mechanics theories for randomly rough surfaces provide a quantitative basis for predicting local pressure and interfacial separation~\cite{Persson2001JCP,Persson2022MRS,muser2017TL}.
They reveal that the relevant physics is multi-scale in nature and excel in predicting leakage of Newtonian fluids~\cite{Dapp2012PRL}.
However, they fall short of describing mixed-lubrication conditions, in particular when large shear rates, high pressures, or flash temperatures affect the lubricant.
Numerical continuum-scale approaches can address many of these complications, including rough-surface lubrication through homogenized mixed-lubrication models~\cite{Scaraggi2011SoftMat}, \hl{coarse-graining approaches to slip-boundary conditions~\cite{Codrignani2023SA}}, smooth-particle hydrodynamics simulations~\cite{Paggi2019Lub}, numerical contact formulations involving elastic~\cite{Putignano2012IJSS} or viscoelastic solids~\cite{Putignano2015IJSS}, and recent multiscale fluid--structure interaction (FSI) frameworks for compliant lubricated interfaces~\cite{Wang2026IJES}. 
Nonetheless, it remains unclear down to what length scales such continuum descriptions remain accurate.

While limited to relatively small scales, 
non-equilibrium molecular dynamics (NEMD) offers a 
molecular-level approach that directly resolves pertinent local dynamics including layering~\cite{Horn1981,Cui2001JCP,Gao2020JCIS} and wall slip~\cite{Bocquet2007SM, Kong2010MSMSE,Codrignani2023SA, Holey2025SA}. 
%
%
Over the last decade, MD has become an increasingly important tool for analyzing friction in elastohydrodynamic~\cite{Lugt2011TT,Ewen2021TL,Ewen2018Fric} and mixed-lubrication~\cite{Spikes1997LS,Stephan2023Fric,Savio2016SA,Eder2011JP} conditions, where experiments are challenging and continuum assumptions can break down. Notably, recent MD studies have reproduced qualitatively the  transition from boundary to mixed to full-film lubrication, i.e., the Stribeck-type behavior~\cite{Stephan2023Fric}.
%
All of these and related studies reviewed in Ref.~\cite{Pastewka2026TI} are restricted to planar or single-asperity geometries.

Motezaker \textit{et al.}~\cite{Motezaker2024APA} extended MD simulations to nanoscale rough contacts and provided a systematic comparison of boundary, mixed, and elastohydrodynamic lubrication regimes.
\hl{Their study convincingly confirmed what had been inferred from macroscopic experiments, showing how lubricant-film thickness, surface roughness, and contact area evolve together across the different lubrication regimes.}
The present work focuses instead on roughness-driven phenomena, using a long simulation cell in the sliding direction to \hl{best reflect random roughness down to the smallest scales} while employing quasi-incommensurate surface alignments to suppress commensurability artifacts, which are known to persist even in lubricated contacts~\cite{Muser2002PRL}.
In addition, we contrast aqueous, hydrocarbon, and immiscible two-fluid lubrication within a common simulation framework.

Lubricants can be broadly categorized into water- and oil-based types. 
Hydrocarbon lubricants have been extensively studied under confinement and shear; they can form robust load-bearing films but also display strong pressure- and shear-dependent rheology and wall slip that depend on molecular structure and surface chemistry~\cite{Restrepo2019behaviour, Mehrnia2023CET,  barsky2001molecular, Kong2009RCS}. 
From an application and sustainability perspective, aqueous lubrication is appealing. However, its performance in metal–metal contacts is often limited by fluid instability, insufficient load-bearing and subsequent reduced surface protection under mixed lubrication, even when full-film friction is very low \cite{yao2026TI}. At the same time, water can exhibit unusual lubrication due to its peculiar structure and dynamics in extreme confinement~\cite{Falk2010NL, Bhamra2024TL}.
%
Interestingly, mixed hydrophilic-hydrophobic systems have been considered rather scarcely, although they hold great promise:
A fluid that is mixed in the bulk phase but phase separates in a contact, where one solid is hydrophilic while the counterface is hydrophobic, is expected to accommodate the shear at a fluid-fluid interface~\cite{deBeer2014NatCom,deBeer2014M}. 

The aim of this work is to contrast the lubrication mechanism of oil-based and aqueous lubricants as well as of their two-fluid immiscible systems for a contact with roughness on several scales, while allowing the confining walls to yield plastically and wall atoms to transfer between the walls. 
To this end, we perform NEMD simulations of the relative sliding motion of two rough copper blocks under a controlled normal compliance, comparing three liquids:
water, $n$-dodecane, and their mixture.
\hl{Water and $n$-dodecane have comparable ambient viscosities (1.0~mPa$\cdot$s for water versus $1.5$~mPa$\cdot$s for $n$-dodecane), but differ in load bearing capacity, confinement response, cavitation behavior, and wall-slip.
The immiscible two-fluid system is expected to add an internal fluid-fluid interface together with opposite wall-fluid affinities.
Its analysis thereby serves as a minimal test of whether interfacial phase separation can reorganize the confined film and suppress asperity engagement as well as material transfer.}
%
Rather than focusing solely on average friction, we analyze time-resolved normal and shear stresses together with configurational diagnostics (cavitation/film rupture, relative slip fields, and post-sliding material transfer) to distinguish event-driven junction dynamics from steady film-supported response. 
Finally, to decouple dissipation mechanisms from trivial load differences, we complement matched-geometry comparisons with pressure-matched simulations via stiffness tuning.

At the system sizes accessible to NEMD, a central difficulty is that matching simultaneously volume fractions and normal pressures across different liquids is infeasible. 
The problem is that the choice of particle number effectively fixes the volume fraction, while the normal pressure emerges from the relative displacement of the confining walls and the coupled solid–fluid response. 
Moreover, at small scales, the response of rough interfaces differs significantly between normal-displacement- and load-controlled conditions, mostly because asperity engagement varies strongly  with lateral displacement.
This partially reflects physical reality.
However, in real systems, such strong spatiotemporal fluctuations are partially compensated by fluid replenishment from neighboring regions, which is not possible in the periodically replicated domains typically used in MD simulations.
Using a grand-canonical technique to fix this issue is not an appealing option, as the systems are inherently far from equilibrium conditions.
Cavitation occurring both in reality and in our simulations would be suppressed by such schemes. 
To nevertheless enable a controlled comparison, we employ an elastic normal coupling of the confining layers (finite stiffness), which interpolates between displacement- and load-controlled limits and allows us to match normal loading conditions across lubricants when desired.  
\hl{In addition, we add some minor, deterministic surface curvature to one of the two confining walls.}

\section{Model and methods}

\subsection{Model}
\subsubsection{Rough copper blocks}
\label{sec:rough_copper}
Two face-centered cubic (fcc) copper blocks are oriented such that their [111] direction aligns with the $z$-axis.
The lattice constant of copper is approximately $a_0 = 3.62$\,\AA, which implies a nearest-neighbor spacing of  
$
r_0 = a_0/\sqrt{2} \approx 2.56~\text{\AA}.
$
This, in turn, defines the lattice constants of an unstrained, orthorhombic unit cell to be
$
a_x = r_0 = 2.56~\text{\AA}, \quad a_y = \sqrt{3} a_x \approx 4.43~\text{\AA}
$, and $a_z \approx 6.27$~\text{\AA} in the standard orientation.

To avoid commensurability artifacts, the bottom block is rotated by 90$^\circ$ about the $z$-axis, which produces the maximum possible in-plane misalignment of $30^\circ$ due to the six-fold symmetry of the fcc (1~1~1) surface.
The top solid remains in the standard orientation.
The ratio of unstrained lattice constants is $\sqrt{3}$, which is about 1\% smaller than 7/4. 
Thus, when applying a compressive strain of about 0.5\% in one direction (the $x$ direction of the top and $y$ direction of the bottom solid) and a tensile strain of 0.5\% in the transverse direction, a square supercell containing $7 \times 4$ elementary cells can be constructed. 
We repeat such a supercell $N_x = 28$ times in the $x$-direction, which yields a total size of $L_x \approx 50$~nm parallel to the sliding direction.
In the normal direction, we consider $N_z=26$ unit cells, yielding an initial height of $L_z = 16.3$~nm.
In the remaining transverse $y$-direction, we merely place one unit cell for reasons of computational efficiency, i.e., $L_y = l_y \approx 17.8$~\AA.
%
Both blocks are initially set up such that the lowest atoms lie in the $xy$-plane.

The topography of the blocks is made randomly self-affine in the long sliding direction, while the system is thin in the transverse direction, which is similar to a recent setup by Bian and Nicola~\cite{Bian2021TI}. 
Specifically, the profile of the interacting surface is defined as:
\begin{equation}
     h(x) = \frac{(x-L_x/2)^2}{2 R_x} + \sum_{n=0}^{n_\text{max}} \sqrt{C(q_n)} \cos(q_n x+\varphi_n)
+ \Delta h,
\end{equation}
where $R_x = L_x$ is used for the indenter so that the slope of the ``deterministic'' height profile is $\pm 1$ at the boundaries of the simulation cell, while $R_x = \infty$ for the substrate,
\sa{while $\Delta h$ is a constant vertical offset used to set the reference height of the generated surface profile.}
The second summand on the right-hand side of the surface profile corresponds to random roughness.
Here, we use 
\begin{equation}
    C(q_n) = \frac{C_0}{\left(1+(q_n/q_r)^2\right)^{(1+2H)/2}}
\end{equation}
with a Hurst exponent of $H = 0.8$, $q_n = 2\pi(n+1)/L_x$, a roll-off wave vector of $q_r = 4q_0$,  and $z_0 = 5 a_0$, which, except for proper normalization, is the height spectrum. 
The phases $\varphi_n$ are uniform random numbers in $[0, 2\pi)$. 
%
%
\hl{The prescribed short wavelength cutoff is $\lambda_s=6a_0$. Since only discrete wavevectors are compatible, a length of $L_x=500~\text{\AA}$ yields 23 Fourier modes, corresponding to a shortest wavelength of approximately $21.74~\text{\AA}$}.
%
%
Finally, the raw profile is shifted downward by $-a_0/2$ and then clipped at zero; hence the final profile used for carving has min($h(x)$) = 0.
This shift and clipping ensure that the outermost asperity does not consist of a single row of atoms. 
After carving, the $z$-positions of the upper wall are inverted and the entire wall is moved up so that its lowest point is initially higher than the highest wall of the substrate.
The precise shift depends on the amount of lubricant immersed between the walls. 

\begin{figure*}[hbt!]
    \centering
    \includegraphics[width=\textwidth]{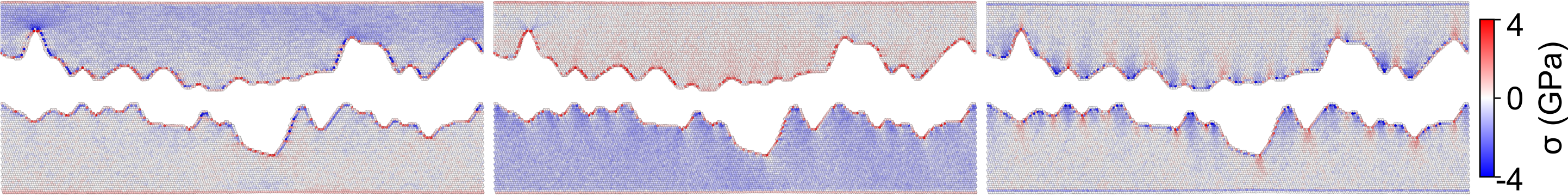}
    \caption{ \label{fig:residual_stress}
    Residual stresses in the simulation blocks after equilibration $\sigma_{xx}$ (left),  $\sigma_{yy}$ (center) and $\sigma_{zz}$ (right). Tensile stress has a positive sign and is indicated in red while compressive is shown as blue. The color range was set to $\pm 4$~GPa for reasons of better visualization. Maximum values for tensile and compressive stress are $4.7$~GPa and $-5.4$~GPa for $\sigma_{xx}$, $4.5$~GPa and $-3.3$~GPa for $\sigma_{yy}$ and $4.3$~GPa and $-5.9$~GPa for $\sigma_{zz}$.
    }
\end{figure*}

After equilibration at $T = 300$~K, unavoidable residual stresses arise in the walls.
As can be seen in Fig.~\ref{fig:residual_stress},  they are largest in outermost layers of both solids.
The compressive stress near surfaces is due to the fact that surface atoms have fewer neighbors than bulk atoms, which makes them ``want'' to shrink their bond lengths; a feature common in metallic and covalent bonds alike~\cite{Muser2022APX}. 

We note that the imposed strain turning a super-cell into a square leads to stress differences between $\sigma_1$ and $\sigma_2$ that results in a bias von Mises stress of approximately 0.49~GPa. This value clearly exceeds 70~MPa, a typical macroscopic yield stress of Copper.
However, our bias von Mises stress does not couple to an active slip plane\hl{, which is why it does not cause plastic deformation.}

\subsubsection{Boundary condition}
\begin{figure}[hbt!]
    \centering
    \includegraphics[width=0.3\textwidth]{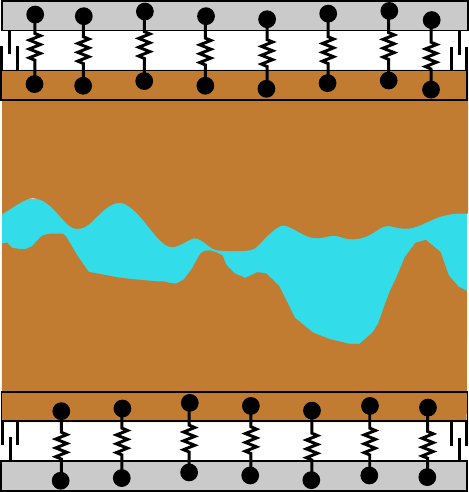}
    \caption{A schematic showing the spring setup used in the present simulations. The atoms in the top and bottom layers are coupled to their initial states via springs in the vertical direction while coupled rigidly in the horizontal direction (as represented by vertical lines).
    \label{fig:spring_setup}
    }
\end{figure}

The choice of boundary conditions—specifically, how the outermost layers are driven—necessarily entails a compromise between controlling the centers of mass (COMs) and allowing physically reasonable internal degrees of freedom (DOFs).
Prescribing COM motion at fixed height yields unrealistically high normal pressures and stress spikes compared with half-space elasticity. Imposing constant normal stress, in contrast, allows excessive height fluctuations, which promotes repeated cavitation and collapse of cavities in elastohydrodynamic-like situations, thereby producing pitting/erosion-like events at artificially high frequency.
The same trade-off applies to internal DOFs: rigid outer layers overestimate elastic restoring forces and amplify stresses, whereas fully unconstrained layers behave like thin plates, underestimating stress fluctuations and again favoring cavitation. Adding many confining layers is computationally inefficient because it would force an unacceptably small cell length in the sliding direction.
Since GFMD is not available in the LAMMPS version used here, we instead couple confining-layer atoms harmonically to ideal lattice sites that translate laterally at constant velocity while being held at fixed height. The spring stiffness 
$k$ interpolates between fixed separation ($k\rightarrow\infty$) and fixed load ($k\rightarrow 0$) and thereby provide a controlled compromise (Fig.~\ref{fig:spring_setup}). The method to determine spring stiffness is described next.

Elastic displacements in (semi-infinite) isotropic solids decay approximately with $e^{-q\Delta z}$, where $\Delta z$ is the distance from the surface.
\hl{The simulated copper block has thickness $\Delta h$, measured from the mean plane of rough contact to the outer boundary; hence the back surface lies at $\Delta z=\Delta h$.}
Although copper is elastically anisotropic, we use this scaling to estimate when the finite slab thickness matters. 
As the mode of wavelength $\lambda$ penetrates \hl{effectively} only to the depth $\sim 1/q=\lambda/(2\pi)$, modes with $\lambda\leq 2\pi \Delta h$ decay well before reaching the opposite slab surface and therefore “see” an effectively semi-infinite substrate; consequently, their stiffness depends only weakly (within a few 10\%) on whether the back surface is treated as zero-stress or zero-strain.
%
%
For a half-space, the normal stiffness per area of such a surface mode is $qE^*/2$.
We therefore choose the spring coupling of the outermost layers to reproduce this half-space stiffness on average over the wavelength range that matters in our finite cell: from the longest accessible mode $\lambda_\text{max}= L_x$
down to the shortest mode that still feels the finite thickness, $\lambda_\text{min} \approx 2\pi \Delta h$. Specifically we match the geometric mean of $k(\lambda_{\text{max}})$ and $k(\lambda_{\text{min}})$, which slightly over-stiffens the longest-wavelength mode by $\sqrt{L_x/(2\pi \Delta h)}$ $(=1.63$ for $L_x=500\text{\AA}$ and $\Delta h=30 \text{\AA}$), an acceptable compromise between overly stiff boundaries and finite-size compliance.
The value for $k$ ultimately turns out to be
\begin{equation}
    k = \frac{\pi E^*\Delta A }{\sqrt{2\pi \Delta h L_x}},
\end{equation}
where $\Delta A =\sqrt{3}r_0^2/4$ is the area per coupled surface atoms for triangular lattice with [111] surface.
This value evaluates to $k \approx 0.35$~N/m using a rough estimate of $E^* = 120$~GPa for the indentation modulus of copper.

\subsubsection{Potential energy surface}
Copper is modeled with an EAM potential \cite{Mishin2001PRB}, while the lubricants are described by TIP4P water \cite{Gravelle2025LJCMS,Zhang2022Nano,Heinz2008JCP} and OPLS-AA dodecane \cite{Jorgensen1996JACS} confined in the interfacial gap. The simulations are carried out for a configuration with maximum penetration ($z^{\text{bot}}_{\text{min}}-z^{\text{top}}_{\text{max}}$) of $\approx2.7$ $\text{\AA}$ with 11,228 water and 917 $n$-dodecane molecules. The parameters used in the simulations are summarized in Table.~\ref{tab:potential_params}. The Lennard-Jones cross-interactions were assigned using Lorentz--Berthelot mixing rules, except for the water--dodecane interactions. For these pairs, the unlike parameters were scaled as
$\epsilon_{ij}=1.07\sqrt{\epsilon_i\epsilon_j},~
\sigma_{ij}=0.85(\sigma_i+\sigma_j)/2$, suggested by Kr\"amer et al.~\cite{kramer2019JCTC}. These modified cross-interactions suppress spurious water--alkane mixing while preserving a stable liquid--liquid interface with a realistic interfacial tension.

\begin{table*}[hbt!]
\caption{Parameters for water \cite{Gravelle2025LJCMS,Zhang2022Nano,Heinz2008JCP} and $n-$dodecane \cite{Jorgensen1996JACS} used in the simulations}
\label{tab:potential_params}
\centering
\resizebox{0.9\textwidth}{!}{%
\begin{tabular}{cccccc}
\hline
\multicolumn{6}{c}{\textbf{Water (TIP4P model)}}                                                                                                                           \\ \hline
\multicolumn{1}{c|}{Atomic properties}             & $m$(amu) & \multicolumn{1}{c|}{$q$(e)} & \multicolumn{1}{c|}{Lennard-Jones parameters}     & $\sigma$(\AA)    & $\epsilon(10^{-3}\text{eV})$    \\ \hline
\multicolumn{1}{c|}{O}                             & $15.999$    & \multicolumn{1}{c|}{$-1.1128$}      & \multicolumn{1}{c|}{O}                            &    $3.1589$       &     $8.03$       \\
\multicolumn{1}{c|}{H}                             & $1.0$   & \multicolumn{1}{c|}{$+0.5564$}      & \multicolumn{1}{c|}{H}                            &   $0.0$        &  $0.0$          \\ \hline
\multicolumn{1}{c|}{Bond potential parameters}     & $K_{\text{bond}}$(eV/\AA$^2$)     & \multicolumn{1}{c|}{$r_0$(\AA)}      & \multicolumn{1}{c|}{Angular potential parameters} & $K_{\text{angle}}$(eV/rad$^2$)         & $\theta_0(^\text{o})$      \\ \hline
\multicolumn{1}{c|}{O-H}                           & 19.514    & \multicolumn{1}{c|}{0.9572}      & \multicolumn{1}{c|}{H-O-H}                        &     2.385      &     104.52       \\ \hline
\multicolumn{6}{c}{\textbf{$n$-Dodecane (OPLS AA model)}}                                                                                                                    \\ \hline
\multicolumn{1}{c|}{Atomic properties}             & $m$(amu) & \multicolumn{1}{c|}{$q$(e)} & \multicolumn{1}{c|}{Lennard-Jones parameters}               & $\sigma$ (\AA)     & $\epsilon(10^{-3}~\text{eV})$    \\ \hline
\multicolumn{1}{c|}{C (CH3)}                       &  12.011   & \multicolumn{1}{c|}{$-0.18$}       & \multicolumn{1}{c|}{C (CH3)}                      &      3.5     &   2.9054         \\
\multicolumn{1}{c|}{C (CH2)}                       &   12.011   & \multicolumn{1}{c|}{$-0.12$}       & \multicolumn{1}{c|}{C (CH2)}                      &      3.5     &     2.9054       \\
\multicolumn{1}{c|}{H}                             &   1.0   & \multicolumn{1}{c|}{$+0.06$}       & \multicolumn{1}{c|}{H}                            &      2.5     &     1.3009       \\ \hline
\multicolumn{1}{c|}{Bond Potential parameters}     & $K_{\text{bond}}$(eV/\AA$^2$)    & \multicolumn{1}{c|}{$r_0$(\AA)}      & \multicolumn{1}{c|}{Angular potential parameters} & $K_{\text{angle}}$(eV/rad$^2$)          & $\theta$($^{\text{o}}$)      \\ \hline
\multicolumn{1}{c|}{C-C}                           &  11.6216    & \multicolumn{1}{c|}{1.529}       & \multicolumn{1}{c|}{C-C-C}                         &  2.7319     &   112.7         \\
\multicolumn{1}{c|}{C-H}                           &  14.7438    & \multicolumn{1}{c|}{1.090}       & \multicolumn{1}{c|}{C-C-H}                         &  1.6262     &   110.7         \\
\multicolumn{1}{c|}{H-H}                           &    -  & \multicolumn{1}{c|}{-
}       & \multicolumn{1}{c|}{H-C-H}               &       1.4310    &    107.8        \\ \hline
\multicolumn{1}{c|}{Dihedral Potential Parameters} & \multicolumn{2}{c|}{$V_1$(eV)}            & \multicolumn{1}{c|}{$V_2$(eV)}                           & \multicolumn{2}{c}{$V_3$(eV)} \\ \hline
\multicolumn{1}{c|}{C-C-C-C}                       & \multicolumn{2}{c|}{0.0754535}              & \multicolumn{1}{c|}{$-0.00680816$}                             & \multicolumn{2}{c}{0.0120986}   \\
\multicolumn{1}{c|}{C-C-C-H}                       & \multicolumn{2}{c|}{0}              & \multicolumn{1}{c|}{0}                             & \multicolumn{2}{c}{0.0158713}   \\
\multicolumn{1}{c|}{H-C-C-C}                       & \multicolumn{2}{c|}{0}              & \multicolumn{1}{c|}{0}                             & \multicolumn{2}{c}{0.0137898}   \\ \hline
\end{tabular}%
}
\end{table*}


\subsection{Methods}

The normal stresses are converted into force per atom in the two outermost layers.
The $x$-positions of the atoms in the outermost layer were constrained to a velocity of $\pm v/2$ in top and bottom wall, respectively.
Their $y$-positions were held fixed at their reference lattice sites, and the $z$-positions were subjected to the harmonic coupling described above.
The next three outermost layer were subjected to a \hl{Grønbech-Jensen} thermostat~\cite{GJ2013MP,GJ2019MP}, \hl{acting} only in the $y$-direction, \hl{perpendicular to both the sliding ($x$) and loading ($z$) directions}. 
\hl{The $x$-component was excluded as it is the direction of shear, while the $z$-component was excluded because thermostatting it would introduce damping and random forces directly into the normal contact dynamics, potentially affecting asperity deformation, squeeze-out, and cavity growth or collapse. Restricting the thermostat to the direction perpendicular to both sliding and loading is therefore a minimally invasive, established strategy in nonequilibrium MD tribology~\cite{deBeer2014Lang,vonGoeldel2021TL,Seidl2021TL}. The thermostat's damping time was set to $\tau = 1$~ps, which is about six times the period corresponding to the Debye frequency of copper.}
The time step was set to $\Delta t = 2$~fs.
All simulations were performed using LAMMPS~\cite{Thompson2022CPC}. 

To visualize the local fluid motion, we estimated a finite-time streamwise velocity from the $x$ displacement of each fluid atom between two consecutively saved configurations.
The displacement field $\Delta x_i$ is computed with respect to the previously saved frame.
The walls move by $d_{\text{span}}=20~\text{\AA}$ between consecutive frames, which defines the averaging span to estimate the velocity.
The normalized streamwise displacement ($x-$direction) of atom $i$ is then calculated as $\tilde{v}_{x,i}=\Delta x_i/d_\text{span}$.
To reduce thermal fluctuations and obtain a spatially resolved flow field is coarse-grained in the $x-z$ plane. The simulation domain was divided into cuboidal bins of size $5~{\text{\AA}}\times17.8~{\text{\AA}} \times5~{\text{\AA}}$. For each bin, the coarse-grained flow field $\tilde{v}_x(x,z)$ calculated by averaging $\tilde{v}_{x,i}$ over the bin. During visualization, the resulting average value is assigned to every particle in the bin.

To suppress thermal noise in the instantaneous per-atom stress, we compute a time-averaged local per-atom virial stress tensor $\boldsymbol{\sigma}_i$ for each copper atom $i$. The time averaging is performed over the aforementioned window. We then form its deviatoric part as~$\mathbf{s}_i = \boldsymbol{\sigma}_i - \mathbf{I}\,\mathrm{tr}\!\left(\boldsymbol{\sigma}_i\right)/3$, and evaluate the von-Mises stress as $\sigma_{\mathrm{vM},i} = \sqrt{(3/2)\,\mathbf{s}_i:\mathbf{s}_i}$.
Similarly, total forces transmitted across the interface are averaged over the same time window to estimate macroscopic normal ($p_{zz}$) and shear ($\tau_{xz}$) stresses. \sa{Because this fixed displacement interval corresponds to different averaging times at different sliding velocities, the resulting fields are used to compare spatial patterns at a common sliding-distance resolution.
The amplitude of fluctuations and the duration of events were not quantified.
}

\section{Results}
Spatially and/or temporally resolved stresses and flow profiles are presented first for individual lubricants, in Sects.~\ref{sec:aqueous}--\ref{sec:mixed}.
Comparisons of {mean} \hl{material-transfer} and friction coefficients between different lubricants follow in Sects.~\ref{sec:wear} and \ref{sec:friction-coefficient}, respectively.
For each lubricant, we investigate relative speeds of 1, 10, and 50~m/s between the two solids.
%
%
\sa{They are intended to probe different local conditions within the same rough mixed-lubrication configuration, ranging from relatively mild to severe conditions.}
A speed of 1~m/s corresponds to low-to-moderate sliding conditions, as encountered in slow-running, highly loaded plain bearings and bushings, as well as related engine bearing contacts~\cite{Scherge2015wear,Baskar2014TI,Rameshkumar2010TI}.
%
%
A velocity of 10~m/s is often reached in gears, where
frictional heating and lubricant entrainment become more important \cite{Liu2020TI}. 
Finally, 50~m/s can be reached in high-speed braking or turbomachinery.
Sliding conditions are severe and similar to those found during grinding or polishing \cite{Wang2025AM}, where local heating, lubricant stability, and direct metal contact strongly affect friction and wear.

\subsection{Aqueous lubrication}
\label{sec:aqueous}

\begin{figure*}[hbt!]
    \centering
    \includegraphics[width=0.95\textwidth]{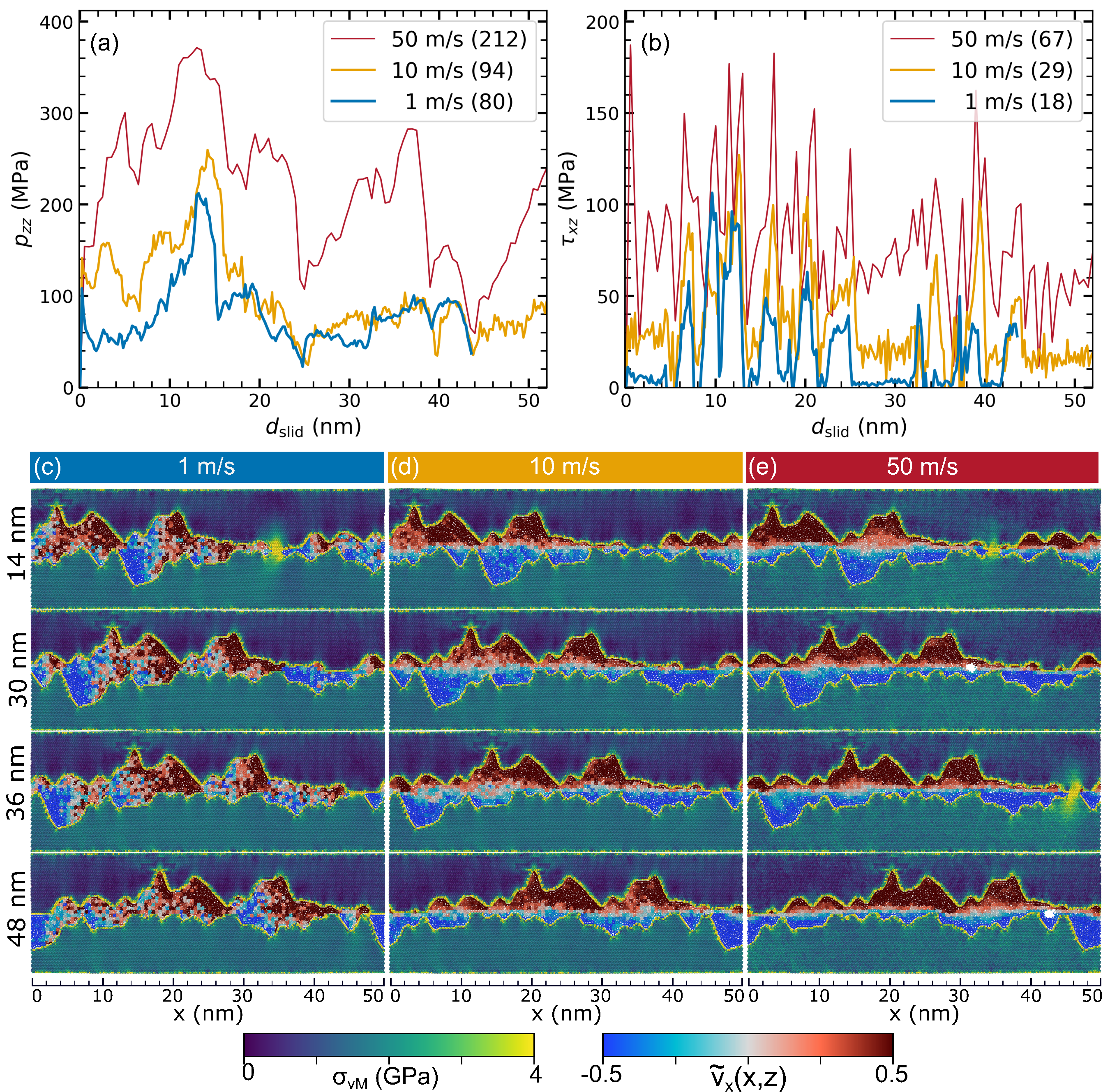}
    \caption{Water-lubricated contact at three sliding velocities. (a) Mean normal pressure $p_{zz}$ and (b) shear stress $\tau$ vs. slid-distance $d_\text{slid}$. Mean values are given in parentheses in the legend.
    Configurational snapshots at four slid-distances are shown for (c) 1 m/s, (d) 10 m/s, and (e) 50 m/s. The corresponding slid-distance ($d_\text{slid}$) is mentioned at the left. The copper blocks are colored by the von-Mises stress, 
    the fluid by the normalized coarse-grained flow field $\tilde{v}_x(x,z)$. 
    \label{fig:water_vel}
    }
\end{figure*}

Fig.~\ref{fig:water_vel} shows the time evolution of the instantaneous but spatially averaged normal pressure and shear stress for the water-lubricated contact at the three investigated sliding velocities, along with configurational snapshots.
They depict the von Mises stress in the confining walls and the flow profiles of the lubricant. 
We first discuss the stresses.

\hl{The bearing pressure increases from $\langle p_{zz}\rangle = 80$~MPa at 1~m/s to 94~MPa at 10~m/s and 212~MPa at 50~m/s, with corresponding mean shear stresses of $\langle \tau_{xz}\rangle = 18$, 29, and 67~MPa. The resulting friction coefficient rises from 0.23 to 0.31 and 0.32, i.e. by roughly 40 \% across the studied velocity range.}
The bearing pressure shows quite large fluctuations with the largest values obtained at the highest velocity.
Fluctuations arise mostly due to different asperity engagement at different moments in time--remember that the employed boundary condition are a compromise between constant normal stress and separation.
The shear stresses change on much shorter time scales than the bearing pressure, for all three velocities.
This can be rationalized by noting that, even under constant-velocity driving, the contact must (quasi-) continuously support the load associated with the relative displacement of the rough walls.
In contrast, shear stress can build up locally and be released abruptly during instabilities, leading to shorter-lived fluctuations.
%
Indeed, each sudden drop in shear stress can be associated with such an instability, for example when shear-shear stress releases abruptly in a direct metal-metal contact or in a highly pressurized boundary lubricated zone. 
The local maxima in the normal pressure occurring at a slid distance of $d_\text{slid} \approx 13 \pm 1$~nm coincides with an asperity engagement taking place near the lateral coordinate $x \approx 34$~nm, as becomes obvious from the bright spot in the von Mises stress, see Fig~\ref{fig:water_vel} panel (e) top row.  
While mean normal and shear stresses at the scale of 50~nm are still relatively small, even at the highest velocity, 
\emph{local} maxima at the moment of asperity collision are quite extreme: 
$p_{zz} \approx 4.9$, 1.9, and 1.5~GPa and
$\tau_{xz} \approx 1.42$, 1.35, and 1.12~GPa,
both times in the order of descending velocities.
These values are of similar order of magnitude as the maximum local von-Mises, 
$\sigma_\text{vM} \approx 5.1$, 3.6, and 2.8~GPa,
occuring during the moment of asperity engagements, see, e.g.,  the panel $d_\textrm{slid} = 36$~nm at $x \approx 45$~nm.
Thus, our mean values for normal pressure is below the macroscopic Vickers hardness of 400~MPa.
However, local values surpass it clearly. 
As already discussed in Sect.~\ref{sec:rough_copper}, the bias von Mises stress of the isolated surfaces surpasses the macroscopically determined values along in-active directions.
In the current simulation, plasticity sets in under sliding when the local von Mises stress is on a similar order of magnitude as found experimentally for single-crystalline whiskers, i.e., up to 6\% of the shear modulus,~\cite{Brenner1956JAP} or 2.65~GPa from DFT-based simulations on \{111\} glide planes~\cite{Roundy1999PRL}.
Our von Mises stresses are slightly larger, because they were not projected onto the easy glide planes.

An interesting feature of the simulations is that large von Mises stresses arise at a slid distance of $d = 14$~nm near $x\approx 36$~nm at the highest and the lowest velocity, but not at $v = 10$~m/s.
This difference appears to be caused by the previous asperity-contact history.
At 10~m/s, a pocket of high pressure lubricant remains trapped under high pressure during asperity engagement. 
This pocket is such that it spreads the supporting-load over a wider area of substrate asperity, thus, protecting it against material transfer.
As a result, material transfer majorly happens from the indenter to the substrate, leading to a smoother local profile before the system reaches ($d_{\text{slid}}=14$)~nm.
%
%

To complete the discussion on stresses in the confining walls, we note that the von-Mises stress in the upper copper block in the bulk differs noticeably from that in the lower block after applying normal forces to the surfaces.
(Note that periodic in-plane strain boundary conditions are applied, which implies that $\sigma_{11} \neq \sigma_{22}$.)
The difference arises because the shear stress $\tau_{xz}$ is finite as well as $C_{14}'$ in the chosen crystal orientation, specifically $C_{14}' = \pm (C_{11}-C_{12}-2C_{44})/(3\sqrt{2})$ for upper and lower wall, respectively.
For the highly elastically anisotropic copper, this evaluates to $C_{14}' \approx -25$~GPa.
Thus, $\sigma_{11}$ and $\sigma_{22}$ differ not only due to the pre-strain making the walls incommensurate but additionally by $2C_{14}'\epsilon_{33}$ with $\epsilon_{33} = \sigma_{33}/C'_{33}$.
(Voigt notation is used for the stiffness tensor and primes indicate the use of tensor elements in the correct coordinate system; stress and strain are retained in tensor form and remain without prime.)
The full rotated stiffness tensor and the resulting stress–strain relations for the unrotated and the 90$^{\circ}$ in \ref{sec:stress_tensor}.
Similarly to the existing von Mises strain before sliding,
the large von Mises stress associated with the imposed strain does not imply that the walls are about to yield: the stress tensor is dominated by normal loading along [111] and has minimal projection onto the FCC {111}⟨110⟩ slip systems. Consequently, it does not effectively activate crystallographic glide despite its large scalar magnitude.

A remarkable feature is the sudden drop in normal pressure near $d_\textrm{slid} = 25$~nm at the highest velocity, see Fig.~\ref{fig:water_vel}(a). 
It coincides with the nucleation of a bubble, visible in panel (e), second row from above.
It also triggers a spike in the friction shear stress, as shown in panel (b).  
While the bubble grows, both normal and shear stress recover to significant parts. 
At the two lower speeds, both the mean pressure and its fluctuations are clearly smaller than at $v = 50$~m/s, and cavitation does not occur.
Moreover, the time evolution of the pressure at the two lower velocities are quite similar for $d_{\text{slid}}>20$~nm, while the shear stress signal for 1~m/s is clearly lower than that of 10~m/s.
This result can be rationalized through an analysis of the lubricants flow profiles, \hl{which follows next}. 

All three flow profiles reveal remarkable deviations from Couette-type flow, which is consistent with the $\lambda$-ratio, defined as $\lambda = h/\sigma$, ranging locally between zero and one.
At the two highest velocities, near-plug flow dominates, with a localized shear zone generally thinner than 1~nm and exhibiting global height variations of about 3~nm.
These small variations are certainly also a consequence of the roughness being negligible in the transverse direction but significant in the longitudinal direction, consistent with solids moving perpendicular to existing wear-track-like features.
%

Flow characteristics are substantially altered at the smallest velocity.
No localized slip plane exists.
Nonetheless, the flow profiles are in clear violation of Couette-type flow even where $\lambda$ is of order unity, indicating that Reynolds-based flow factor approaches are not applicable under these conditions.

At sliding velocities of 10~m/s with shear being localized to roughly 1~nm, the shear stress is expected to be of order 10~MPa.
Thus a non-negligible fraction of the average value of 29~MPa arises from the (approximately Newtonian) shearing of the fluid.
We associate the remaining 19~MPa to instabilities occurring in the solid, such as plastic deformation, having a much weaker rate-dependence than Newtonian flow. 
Assuming that the 10~MPa viscous contribution quintuples when $v$ increases from 10~m/s to 50~m/s, while the remaining $\sim$19~MPa  remains approximately unaltered, yields a shear stress of 69~MPa at the highest velocity.
This is close to the measured 67~MPa.
The agreement with the measured value certainly benefits from fortuitous error cancellation.
For example, on one hand, even water is expected to shear thin at extremely high shear rates or its viscosity to decrease due to local heating. 
On the other hand, additional dissipation mechanisms such as cavitation contributes at $v = 50$~m/s but not at $v = 10$~m/s (at least not at our system sizes).
\sa{At $v=1$~m/s, the flow morphology changes substantially: the shear is less localized and the shear-zone becomes more strongly correlated with the local orientation of the rough surfaces.
Translation of an inclined asperity flank locally opens or closes the available fluid volume.
In opening regions, the liquid can remain largely entrained by the adjacent wall, whereas closing regions require lateral redistribution of the confined liquid and therefore generate stronger relative motion.
}
At $v = 1$~m/s, the viscous contribution to the shear stress becomes small, since the velocity is reduced and the shear zone substantially broadened compared to the intermediate velocity.
The friction is then likely dominated by processes in the confining walls, where numbers match up again surprisingly well, i.e., 18 vs. 19 MPa. 
Although all estimates are certainly rough and hence must be taken with a grain of salt, we believe that the presented arguments account for the observed trends in leading order. 
%

%

%

\subsection{Non-aqueous lubrication}
\label{sec:non-aqueous}

\begin{figure*}[hbt!]
    \centering
    \includegraphics[width=0.99\textwidth]{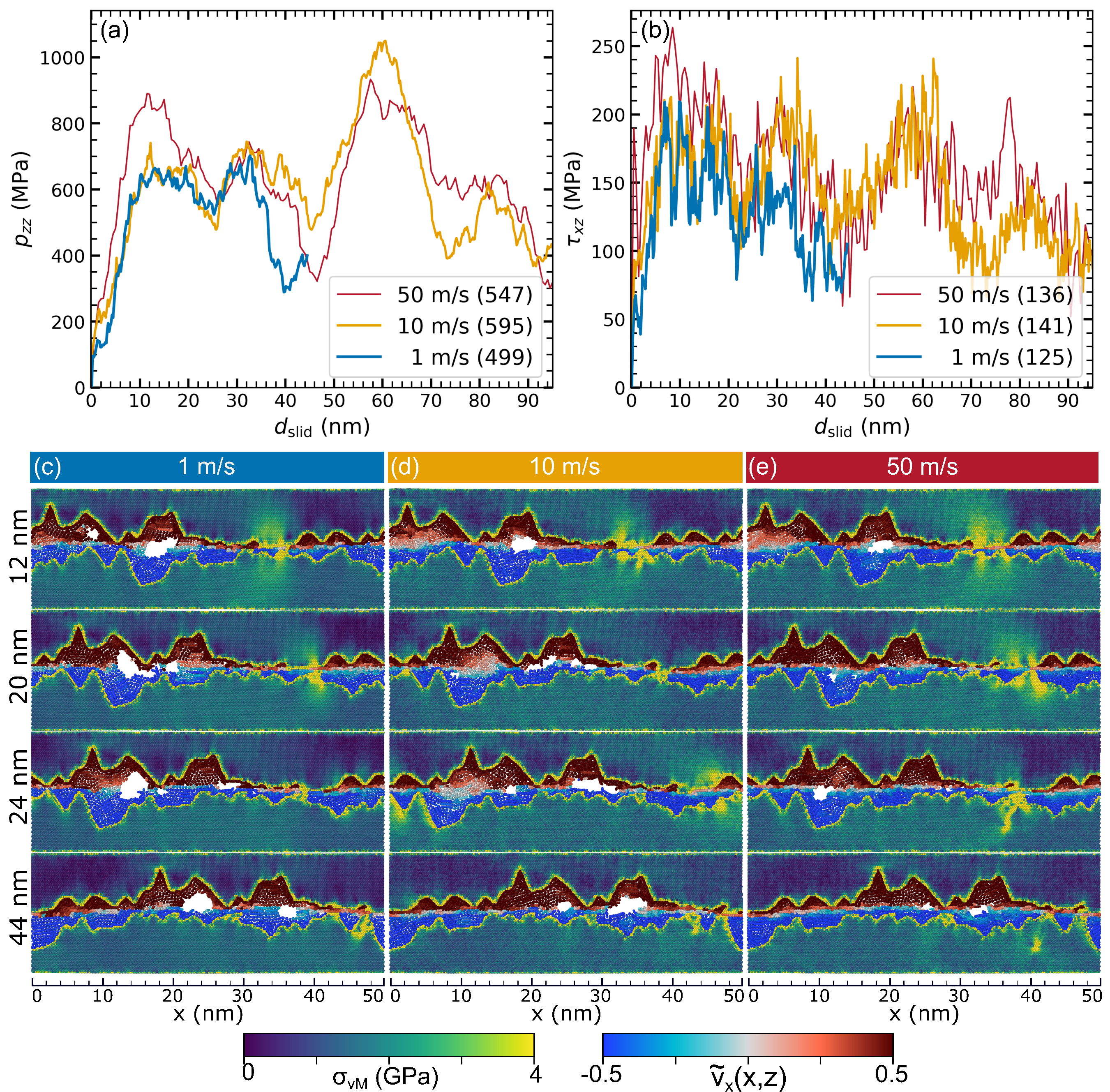}
    \caption{
    Same as Fig.~\ref{fig:water_vel}, but for $n$-dodecane.
    \label{fig:dod_vel}
    }
\end{figure*}

Figure~\ref{fig:dod_vel} summarizes the normal pressure, shear stress, and flow morphology for the $n$-dodecane-lubricated contact. 
All three quantities show a rather weak sensitivity to the sliding speed over the range studied, in particular in comparison to the water film. 
Under the matched initial filled volume, the mean normal pressure remains high at all three velocities, with $\langle p_{zz} \rangle \approx 499$--$595$ MPa.
This indicates  a better load support compared to water, since a greater load is carried at the same initial separation and volume filling. 
The modest variations in mean stress with velocity, $\langle \tau_{xz} \rangle \approx 125$--$141$ MPa, imply  a nearly velocity-independent friction coefficient, i.e., $\mu \approx 0.24$--$0.26$.

As compared to water, more local von Mises stress 
maxima occur.
They are reflected as crests in the $p_{zz}$ response in Fig.~\ref{fig:dod_vel}(a), e.g., two small peaks on the first global maximum of the 50~m/s curve near a sliding distance of $d_\text{slid}\approx 12$~nm. 
These local maxima are associated with asperity approach during which lubricant remains trapped between the opposing surfaces. As the asperities move closer, the confined hydrocarbon film is compressed and develops a squeeze-film-like load-supporting pressure. This local pressure can either separate the surfaces and delay direct metal--metal contact or yield under the mixed stress-separation boundary conditions.
When the lubricant is expelled completely, the mean normal pressure does not always spike (see $d=24$~nm for 1 and 10~m/s in Fig.~\ref{fig:dod_vel}(c,d)), although the von Mises stress has 
a  prominent local maximum.
Such maxima occur simultaneously with large (mean) shear stresses, which suggests that tangential forces are transmitted through adhesive contact, in agreement with the most classical explanation summarized by Bowden and Tabor~\cite{Bowden2001book}.
As a consequence of the described dynamics, pressure and shear do not necessarily peak simultaneously.

Despite the normal pressure being large, the $n$-dodecane lubricated contact retains one to three cavities, after the first cavity is nucleated after about 4~nm of sliding.
The cavity is not a single stationary void. Instead, it continuously evolves: it stretches, deforms, fragments into smaller voids, collapses locally, and new ones appear at other locations as the rough surfaces slide past each other.
Compared to water, cavities occur much more frequently.
This increased frequency can be rationalized by the lower nucleation barrier, $\Delta G \propto \gamma^3/\Delta p^2$, where $\Delta p$ is the tensile hydrodynamic pressure generated in the gap, which can be estimated to lowest order from the Reynolds equation.
However, their traces in the friction are less visible.
This is because capillary stresses, which scale as $\gamma/R$, are smaller in $n$-dodecane due to its lower $\gamma$ and larger cavity size $R$. Moreover, the characteristic time scale of the cavity dynamics, $\tau \propto \eta R/\gamma$, is significantly larger in $n$-dodecane.
As a result, the cavity-caused dissipation in $n$-dodecane is distributed over longer times and does not produce sharp features in the friction signal.

The flow maps show that high-slip regions remain confined to a narrow interfacial band and barely expand at lower speed. In fact, slip occurs over such short distances even at 1~m/s to the extent that a continuum description becomes inapplicable.
In particular, it is not meaningful to describe the situation in terms of slip lengths or a local Reynolds equation.

\subsection{\hl{Two-fluid} lubrication }
\label{sec:mixed}

In the \hl{two-fluid}-lubrication setup, we include both fluids, using half the number of water molecules and half the number of $n$-dodecane molecules compared with the corresponding single-fluid simulations. To impose preferential wall-fluid affinities, we tune the wall–fluid Lennard–Jones energy parameter $\epsilon$ such that the top copper wall \hl{preferentially attracts water, whereas the bottom preferentially attracts $n$-dodecane.} Specifically, we double $\epsilon$ for the top–water and bottom–dodecane interactions, and halve $\epsilon$ for the complementary pairs (top–dodecane and bottom–water), thereby biasing water to the top surface and dodecane to the bottom surface. 
\hl{These interaction strengths are intended to impose a qualitative contrast in wall affinity rather than to reproduce a specific experimental contact angle.}

\begin{figure*}[hbt!]
    \centering
    \includegraphics[width=0.99\textwidth]{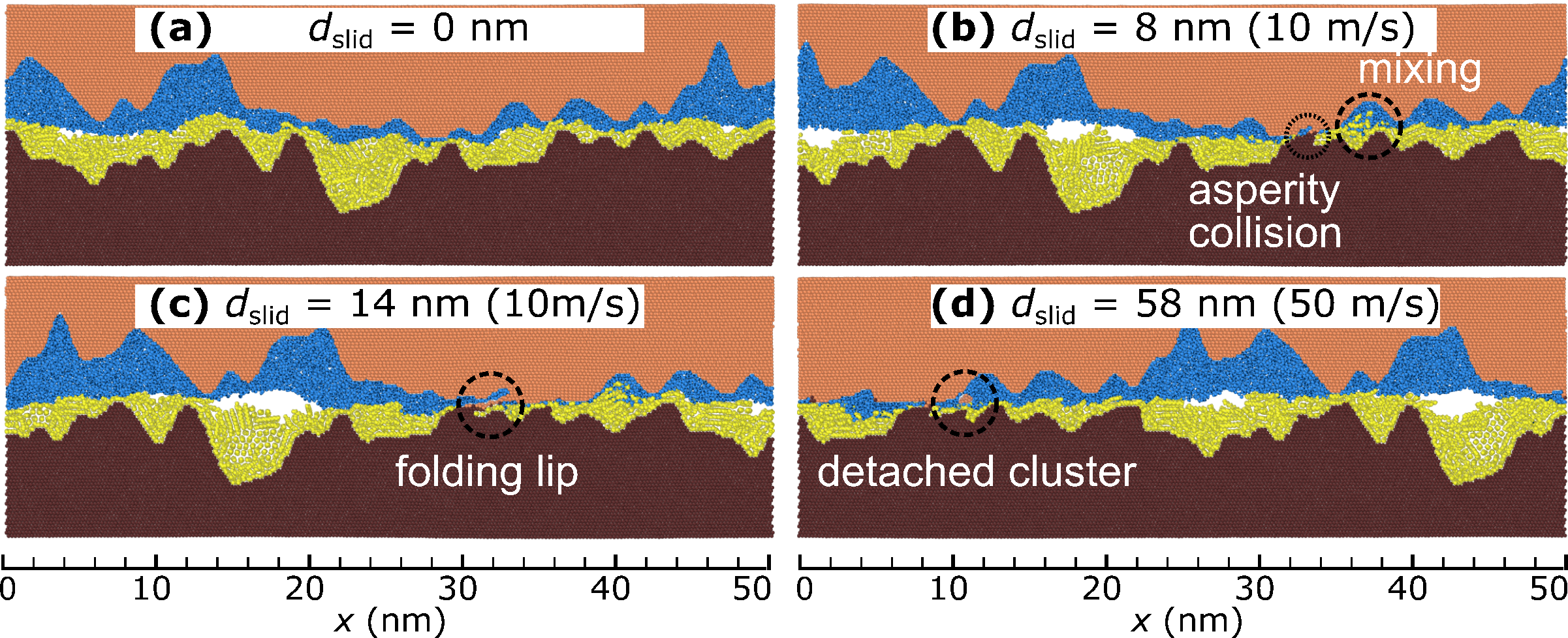}
    \caption{Interface evolution for \hl{two-fluid} lubricated contact. Water is represented in blue near the top wall while $n$-dodecane in yellow near the bottom wall. The top wall moves towards the right while the bottom wall moves towards the left.
    \label{fig:mixed_vis}
    }
\end{figure*}

Before presenting the stress and flow profiles, we discuss a few distinctive features of the \hl{two-fluid} system.
As shown in Fig.~\ref{fig:mixed_vis}(a--d), the two liquids remain phase separated for most of the simulation, except for short intervals following asperity collisions.
Cavities preferentially extend along the fluid--fluid interface and are therefore considerably more elongated than in the single-component systems, as shown in Fig.~\ref{fig:mixed_vis}(b) at $x\approx38$~nm.
Material transfer proceeds differently as well: elongated lips emerge after asperity collisions, shifting the associated dynamics to substantially larger time scales (Fig.~\ref{fig:mixed_vis}(c) at $x\approx32$~nm).
At sliding speeds of 10~m/s or lower, material in these lips has sufficient time to be either reincorporated into the originating surface or transferred locally to the opposing surface.
%
At the high sliding speed of 50~m/s, however, such overhangs can instead detach and form \hl{isolated clusters}, as shown in Fig.~\ref{fig:mixed_vis}(d) near ($x\approx12$)~nm. 

\hl{The illustrated cluster originates from the water-preferential surface and remains within the water-rich phase after detachment. As the sliding proceeds, it encounters another nearby water-preferential asperity of the same block and reattaches. Although the cluster remains close to the liquid-liquid interface, it does not migrate to $n$-dodecane rich region due to its preference towards water. 
Both folding-lip formation and the transient local mixing following strong asperity engagement recur for an independently generated roughness realization (\ref{sec:rough_real_2}), indicating that they are not tied to the particular asperity arrangement studied here.}

\begin{figure*}[hbt!]
    \centering
    \includegraphics[width=0.99\textwidth]{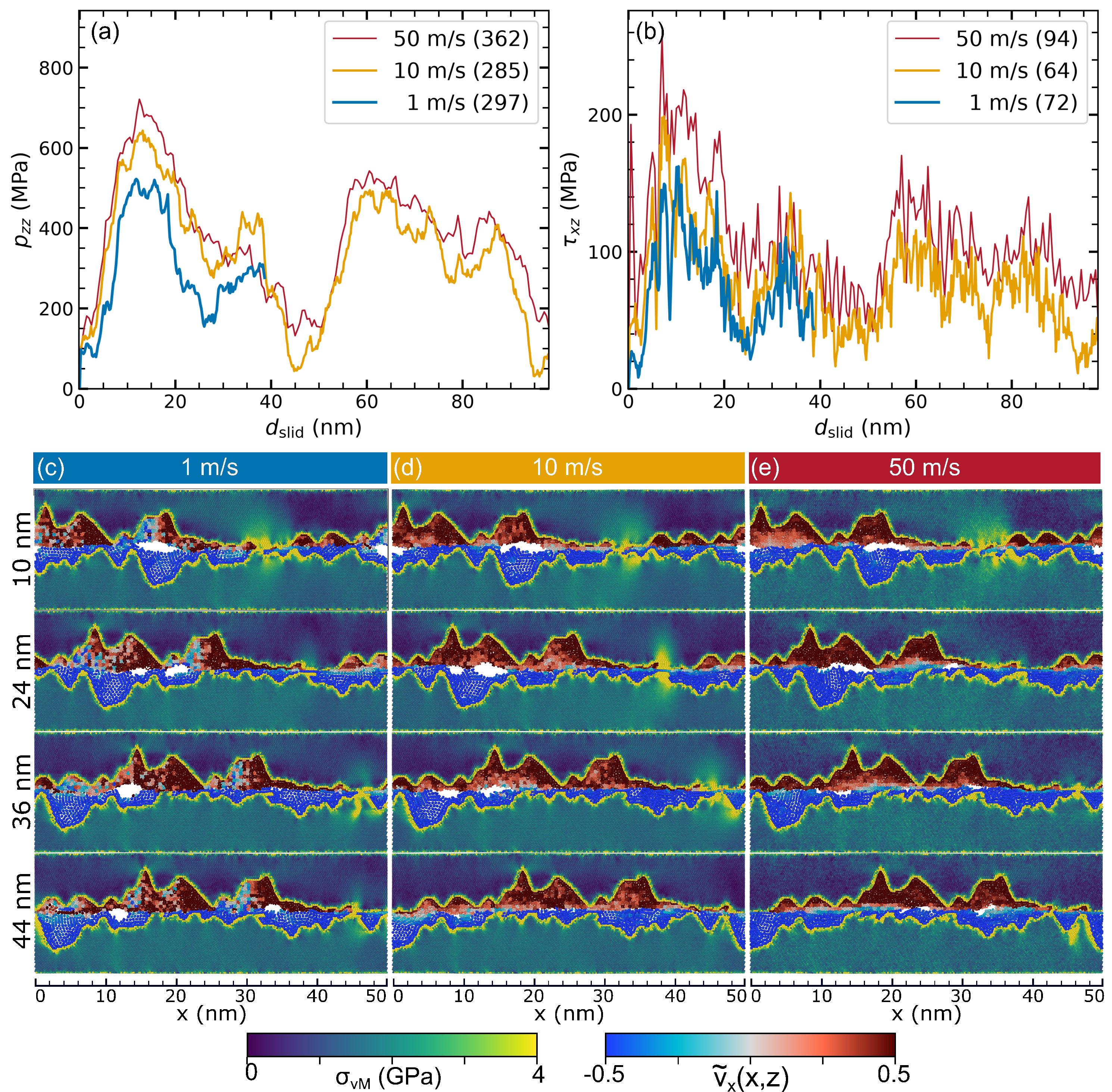}
    \caption{
    Same as Fig.~\ref{fig:water_vel}, but for \hl{two-fluid} lubricated contact. 
    \label{fig:mixed_vel}
    }
\end{figure*}

Under the given number of total molecules and fixed separation of the reference layers, the time dependence of the normal and shear stresses are in between those of water and $n$-dodecane, though clearly closer to the latter, as can be seen by comparing the panels (a) and (b) of Fig.~\ref{fig:mixed_vel} with the corresponding panels of Fig.~\ref{fig:water_vel} and Fig.~\ref{fig:dod_vel}.
One distinguishing feature is that the friction at the smallest velocity is clearly reduced to that of the $n$-dodecane system.
This is consistent with the corresponding flow field at 1~m/s: \hl{water remains mobile and is more strongly redistributed by roughness, whereas $n$-dodecane shows a more persistent plug-like response.
The two-fluid system thus exhibits an intermediate behavior, retaining a partial plug-like flow at 1~m/s; compare Fig.~\ref{fig:mixed_vel}(c)  with Figs.~\ref{fig:water_vel}(c) and \ref{fig:dod_vel}(c).
%
}
%

\subsection{Comparison of material transfer}
\label{sec:wear} 

\begin{figure*}[hbt!]
    \centering
    \includegraphics[width=0.99\textwidth]{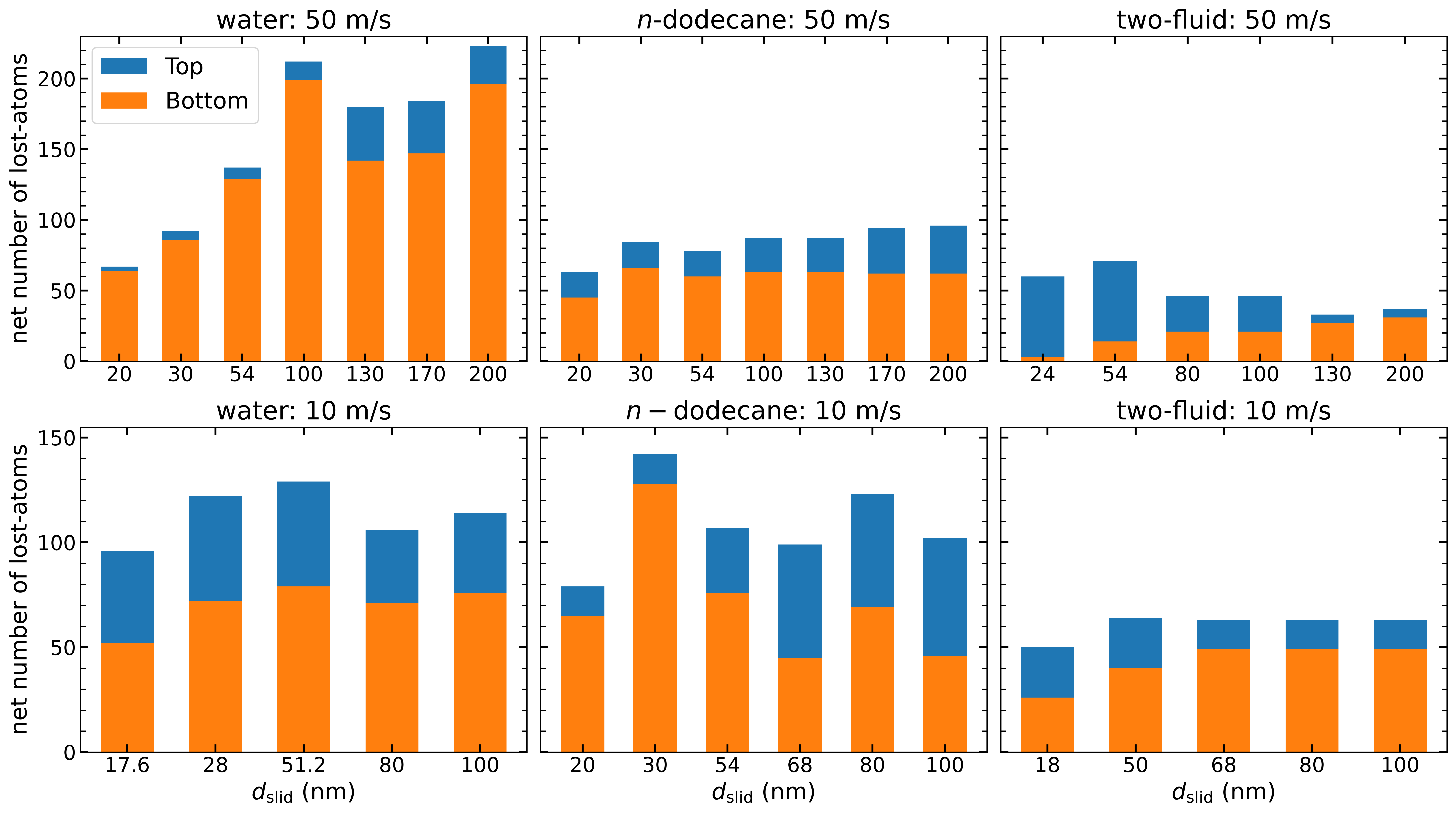}
    \caption{Number of atoms transferred from one wall to the other as a function of sliding distance $d_\text{slid}$ for different sliding speeds, i.e., 50~m/s (top row) and 10~m/s (bottom row), and different lubricants, i.e., water (left), $n$-dodecane (center), and the \hl{two-fluid lubricant} (right). 
     ``Top" gives the number of atoms transferred from the top to the bottom-wall, and vice-versa for ``Bottom".
    \label{fig:wear}
    }
\end{figure*}

In all simulations, 
\hl{but for the low-velocity, pressure-matched case of $n$-dodecane, which is reported further below in Sect.~\ref{sec:pressure_match},}
atoms originally belonging to one solid transfer to the other.
Fig.~\ref{fig:wear} summarizes the respective cumulative numbers, separated by the direction of transfer (``Top'': atoms removed from the top block and incorporated into the bottom; ``Bottom'' vice versa). Three main trends emerge:
First, the transfer is directional: more atoms transfer from bottom to top than the other way around except at the onset of sliding in the \hl{two-fluid} 50~m/s system.
This can be attributed to the imposed strain:
The top surface is slightly compressed in the sliding direction, making this its stiff direction, while the bottom surface is softened in that direction. 
\hl{
Another reason is that the upper wall is slightly rougher, because a deterministic roughness profile was added to it. 
As material is lost predominantly to eliminate surface roughness, preferential wear is observed on the more strongly curved surface, as is corroborated in more detail in~\ref{sec:app_swap_wall}.
}
Second, the counts of \hl{transferred atoms} are not strictly monotonic with sliding distance. 
Occasional reductions reflect reciprocal mass exchange: atoms that are first transferred from, e.g., the top  to the bottom block may later transfer back at a different location. 
Third, \hl{material-transfer} depends strongly on the lubricant and in the case of water also on speed:
at 50 m/s, water produces the largest material transport and generally increasing \hl{atom-transferred-}counts with sliding distance, whereas $n$-dodecane and the \hl{two-fluid} cases yield substantially lower totals over the same range, which is similar for 10~m/s and 50~m/s.  
At all velocities, the \hl{two-fluid} case shows the clearest suppression of material transfer relative to the pure lubricants.

The second and third findings could be seen as a violation of the Reye-Archard law~\cite{Reye1860CI,Archard1953JAP}, which states that wear is proportional to load and sliding distance but insensitive to velocity. 
However, material transfer is not to be equated with wear. More importantly, the system is too small to (a) produce wear particles and, crucially, (b) sample over many individual asperity collisions, as they occur in macroscopic systems.
\hl{
Nevertheless, such local junction and material-transfer events can contribute to the progressive evolution of surface topography during sliding, and in fact, elongated copper-copper transfer has recently been identified experimentally~\cite{Xu2026arx}.
}
Finally, when normalizing the transfer counts by the normal load, we obtain—after a sliding distance of approximately 100~nm—0.94 and 1.17~atoms/MPa for water at 50~m/s and 10~m/s, respectively; similarly, we find 0.17 and 0.17~atoms/MPa for $n$-dodecane, and 0.14 and 0.18~atoms/MPa for the \hl{two-fluid} case.
Thus, the transfer counts normalized by normal pressure show only a marginal velocity dependence. 
Moreover, the \hl{immiscible two-fluid} lubricant only provides a small overall \hl{material transfer} reduction compared to $n$-dodecane. 

\subsection{Comparison of friction-coefficient}
\label{sec:friction-coefficient}

\begin{figure}[hbt!]
    \centering
    \includegraphics[width=0.5\textwidth]{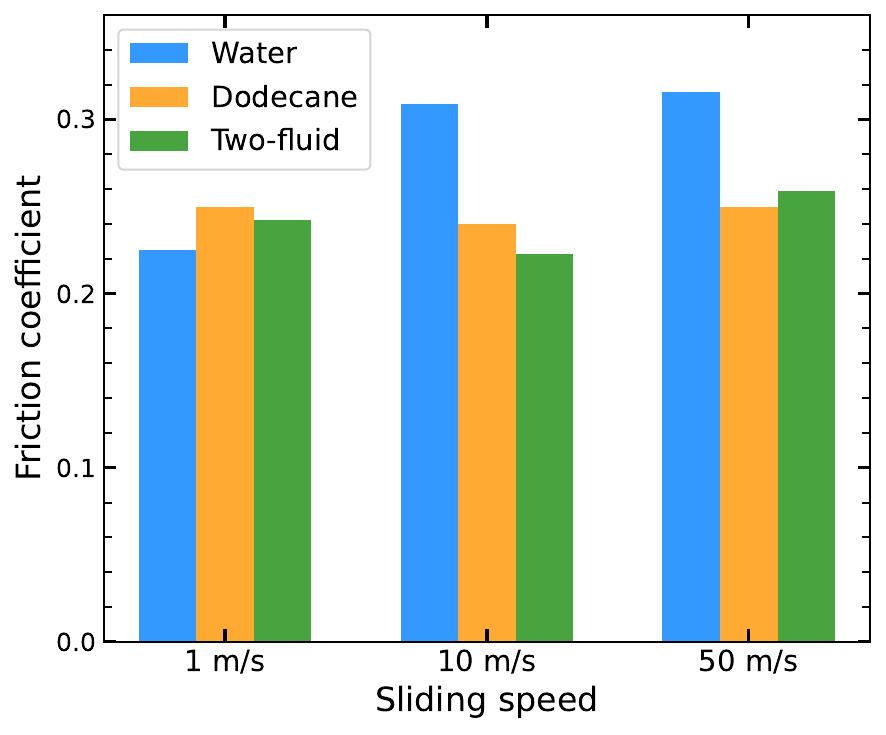}
    \caption{Effect of lubricant and sliding speed on friction coefficient.
    \label{fig:friction}
    }
\end{figure}

In our simulations, the three lubricants exhibit qualitatively different film morphologies and stress responses. Figure~\ref{fig:friction} compares the mean friction coefficient $\mu$ as a function of sliding speed for water, $n-$dodecane, and the \hl{two-fluid} lubricant. 
The contact lubricated with $n$-dodecane exhibits a nearly speed-independent response ($\mu \approx$0.24–0.25), indicating a robust and stable shear–load partitioning over 1–50 m/s.
Water behaves differently: $\mu$ is speed sensitive, showing initial steep rise from $\approx 0.23$ at 1 m/s to $\approx0.31$ at 10 m/s, followed by a non-significant rise to $\approx0.32$ at 50 m/s.
It implies a speed-dependent ranking—water performs better than $n-$dodecane at low speed but becomes less favorable at intermediate and high speeds. Moreover, cavitation and exacerbated \hl{material-transfer} makes its usage at high speed unfavourable.
%
%
\hl{The two-fluid lubricant tracks $n$-dodecane at 1 and 50~m/s ($\mu=0.24$ and 0.26 versus 0.25 and 0.25) and is the lowest of the three only at 10~m/s (0.22 versus 0.24 and 0.31).}
Even at 10 m/s, the reductions in friction remain modest compared to polymer-brush systems, where both can be suppressed by more than an order of magnitude~\cite{deBeer2014NatCom,deBeer2014M}.


\subsection{Comparison under approximately matched pressure}
\label{sec:pressure_match}

So far, we have used the same spring stiffness $k$, approximately the same initial mean separation, and the same lubricant volume in all cases studied. Under these conditions, the differences in normal pressure arise from differences in the load-bearing capacity and the ability to conform to the rough walls.
To decouple the friction response from this pressure difference, we perform the analysis under \hl{approximately} matched pressure conditions. 
In principle, the target pressure could be reached either by increasing the mean gap or by reducing the normal stiffness. 
Increasing the mean gap, however, would suppress asperity interactions and thus undermine a meaningful assessment of material transfer. 
We therefore adjust the spring stiffness $k$ in the $n-$dodecane and \hl{two-fluid} simulations to bring $\langle p_{zz}\rangle$ into \hl{an approximate} agreement with the water case.
\hl{The stiffness was kept at $k=0.35$~N/m for water, but reduced to $0.023 k$ for $n$-dodecane, and $0.055 k$ for the two-fluid case.}
\hl{This procedure necessarily also changes the compliance of the confining boundaries and may therefore influence stress fluctuations and asperity engagement.
Thus, the matched-pressure simulations should not be interpreted as a comparison under otherwise identical mechanical boundary conditions, but rather as a complementary analysis in which the lubricant-dependent differences in mean normal pressure are substantially reduced while approximately preserving the geometric confinement and lubricant volume.}

\begin{figure*}[hbt!]
    \centering
    \includegraphics[width=0.9\textwidth]{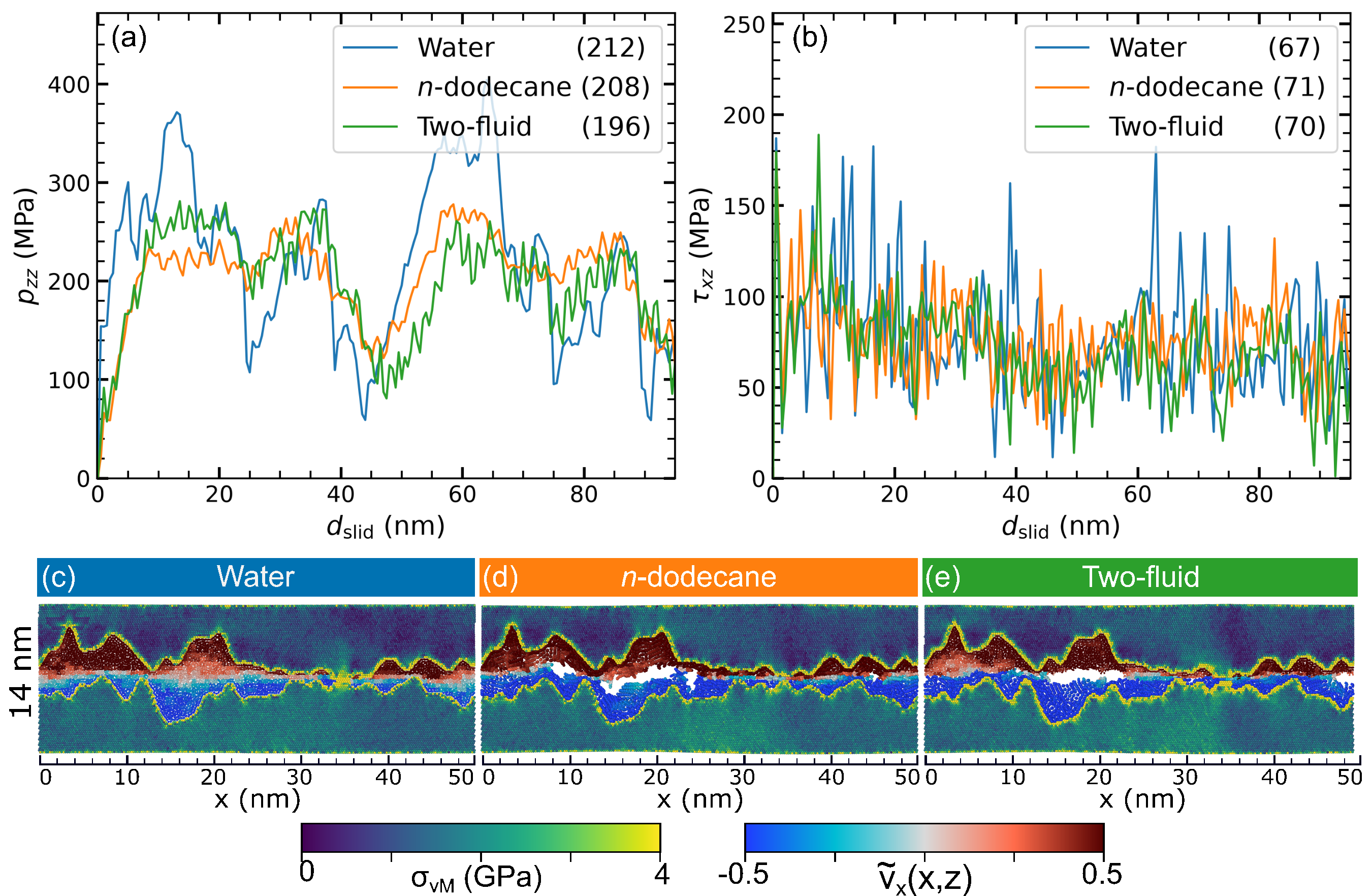}
    \caption{
    Similar to Fig.~\ref{fig:water_vel}, but for three lubricants at one velocity at pressure-matched conditions.  
    The ratio of mean traction and mean normal load are $\mu = 0.32$, 0.34, and 0.36 for water, $n$-dodecane, and \hl{two-fluid}, respectively. Configurational snapshots after slid-distance of 14~nm are shown for (c) water, (d) $n$-doecane, and (e) \hl{two-fluid} case. 
    The color coding is the same as in Fig.~\ref{fig:water_vel}.
    \label{fig:friction_soft}
    }
    \end{figure*}

Figure \ref{fig:friction_soft} compares the three lubricants at 50~m/s under  pressure-matched conditions, i.e., approximately the same mean pressure as for water.
The pressure and shear-stress variation is shown in \ref{fig:friction_soft}(a,b) respectively. 
As can be seen the three lubricants now yield similar shear stresses, resulting in friction coefficient of 0.32 for water, 0.34 for $n$-dodecane, and 0.36 for the \hl{two-fluid} case.
Another important difference appears in the local contact morphology, as shown in local configurations in \ref{fig:friction_soft}(c,d,e).
In the water-lubricated case, the full squeeze out, thus the asperity engagement, happens often.
In contrast, for $n$-dodecane asperity contact never happens during the simulation, indicating that the hydrocarbon film provides the most effective protection against direct asperity engagement.
The immiscible \hl{two-fluid} lubricant lies between these two limits.
The two liquids remain largely phase separated during sliding.
Direct or near-direct asperity contacts occur occasionally, although less persistently than in water.

\hl{It is worth noting that the water and $n$-dodecane do not have comparable viscosities at 200~MPa: the viscosity of $n$-dodecane exceeds that of water by roughly an order of magnitude.
Nevertheless, the friction coefficients for both agree to within 7\%.
Bulk viscosity therefore does not set the level of friction here.
The imposed motion is accommodated by slip in narrow interfacial bands rather than by viscous shearing of the full film. The corresponding local shear rates place the $n$-dodecane film well beyond its Newtonian range, so the transmitted stress is far smaller than a bulk-viscosity estimate would suggest. In the aqueous case, a comparable stress is reached through a different route, since part of it originates in the frequent asperity engagements and the associated plastic deformation. Similar friction coefficients thus arise from dissimilar mechanisms, which reinforces that friction and surface protection should be interpreted separately.}

\subsection{Local thermal response}

The preceding sections show that the friction response is controlled by localized, intermittent events rather than by steady Couette-like shear. 
In particular, asperity engagement, confined-film squeeze-out, plug-like flow, and cavitation collapse produce abrupt changes in the normal and shear stresses. 
To identify where the corresponding mechanical work is dissipated, we examine local kinetic-temperature fields for the most severe water-lubricated case. 
The purpose of this analysis is not to validate a continuum flash-temperature model, but to provide a spatial diagnostic of \hl{where instabilities and thus dissipation occur, which is done in a similar spirit as in previous work~\cite{Gao2021L}}.
Local temperatures are determined using a profile-unbiasing estimator, which subtracts the center-of-mass velocity along the sliding direction of all atoms in  bins of dimension $5~\text{\AA} \times 17.8~\text{\AA} \times 5~\text{\AA}$. 
The local temperature $T_k$ is then determined using equipartition theorem:~$T_k = \sum_{i\in\text{bin }k} m_i \left| \mathbf{v}'_i \right|^2/(g_k N_k -1)k_\text{B}$, where $m_i$ is the atomic mass, $k_\text{B}$ is the Boltzmann constant, and $g_k$ represents the degrees of freedom per atom ($= 2$ for the constrained water molecules and $= 3$ for the unconstrained solid phase). 
These bin-localized values were temporally averaged over 2,000 time steps and projected back to the individual constituent particles ($T_i = \langle T_k \rangle$) to construct the continuous spatial field. 
For visualization, temperature profiles are spatially smoothened using a particle-wise Gaussian filter ($\sigma = 2.0~\text{\AA}$, $r_{\text{cutoff}} = 5.0~\text{\AA}$) via a $k$-d tree neighbor search \cite{Bentley1975CommACM}  to eliminate high-frequency statistical noise.

\begin{figure}[hbt!]
    \centering
    \includegraphics[width=0.95\textwidth]{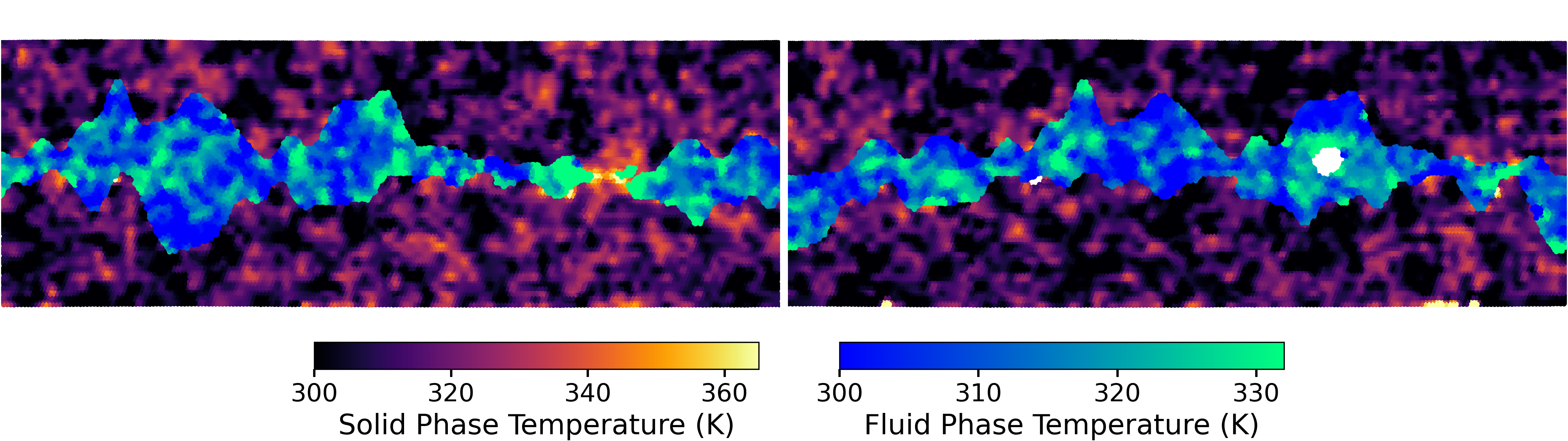}
    \caption{Local temperature maps for solid and fluid (water) phase during (left) asperity engagement and (right) cavitation collapse at 50 m/s sliding speed.
    \label{fig:local_temp_water}
    }
    \end{figure}

Fig.~\ref{fig:local_temp_water} shows the local temperature fields in the solid and liquid phase for water-lubricated sliding at 50~m/s sliding speed.
They exhibit spatial heterogeneity, with warmer regions near asperity engagement (left panel of Fig.~\ref{fig:local_temp_water}) 
and during cavity collapse (right of Fig.~\ref{fig:local_temp_water}).
Using the transient estimate~\cite{Muser2026PRB}, $T_0(t_c)=2\dot{q}\kappa^{-1}\sqrt{Dt_c/\pi}$, with $\dot{q}=\chi\tau_{\text{loc}}v$, $\chi=1/2$, and contact time $t_c=l_c/v$,
a \textit{local} shear stress of 1~GPa over an 8~nm contact length gives \hl{an estimate of} $\Delta T\approx10$~K for copper ($D=1.16\times10^{-4}$~m$^2$/s, $\kappa=400$~Wm$^{-1}$K) at $v=50$~m/s.
%
\hl{The relative change of surface temperature is not affected significantly by how far the thermostat is placed away from the wall, as soon as the slab height reaches the length of the simulation cell over $2\pi$~\cite{Muser2026PRB}. 
This is why the heated zones in our simulations do not extend to the vicinity of where the thermostat acts. 
Placing the thermostat further away from the interface would predominantly increase the base temperature~\cite{Muser2026PRB}.
}
Future work will address the sensitivity of the temperature field to this modeling choice.

%
%


\section{Conclusion}

In this work we use non-equilibrium molecular-dynamics to study lubricated sliding between rough, deformable, quasi-incommensurate metallic contacts.
To probe the different stress-transfer mechanisms, we chose three lubrication cases: aqueous (water), hydrocarbon ($n$-dodecane), and their immiscible mixture. 
Since water and $n$-dodecane have comparable ambient viscosities, their different behavior reflects differences in film stability, load-bearing capacity, wall slip, and confinement response rather than viscosity alone.
The immiscible mixture adds an internal fluid–fluid interface and preferential wall-fluid affinities, allowing us to test whether two-fluid morphology can suppress asperity engagement.

\hl{A key finding is the pronounced departure from Couette flow.
For most of the simulations, shear is often localizded in a band thinner than 2~nm, and the flow is close to plug-like elsewhere.
Over such length-scales a description in terms of slip-length or a local Reynolds equation is no longer meaningful.
A Couette-like velocity profile should therefore not be assumed a priori for rough lubricated contact at nanometric separations.
These intense local shear gradients significantly lower the threshold for cavitation, allowing it to occur at surprisingly moderate average shear rates.
}

\hl{The familiar mixed-lubrication pathways remain clearly identifiable even at these scales: a pressurized confined film, film rupture or cavitation, and direct or near-direct asperity engagement.
The significance of the present work does not lie in identifying these already established mechanisms, but in showing how they compete and interchange over nanometric distances and couple to roughness, stress-distribution, and material-transfer.
Cavitation can be triggered by asperity encounters and release shear stress over a region much larger than the cavity itself, whereas direct asperity engagement promotes plastic deformation and material exchange.
Lubricant chemistry strongly affects the persistence of these states: water more readily loses fluid-borne support, whereas $n$-dodecane maintains a more persistent confined film.
Mechanisms familiar from macroscopic mixed lubrication thus remain physically meaningful at these scales, but their local manifestation is highly intermittent and strongly coupled to the evolving roughness.
}

\hl{The two-fluid system adds another level of mechanical reorganization.
Preferential wall–fluid affinities combined with liquid–liquid immiscibility reorganize the confined film, modify the local separation, and affect plastic deformation and material transfer between the walls.
The local wall profiles therefore evolve more rapidly than in the single-lubricant cases, reducing subsequent asperity interaction and suppressing further material transfer.
Although the liquid–liquid interface can provide an additional slip plane, 
this does not necessarily translate into reduced friction under rough confinement.
Accordingly, the two-fluid lubricant provides no universal friction advantage: its friction is lowest only at 10 m/s under common-stiffness conditions, while no benefit is observed in the approximately pressure-matched 50 m/s case. 
An independent roughness realization reproduces the principal mechanisms while showing that their quantitative manifestation depends sensitively on the local confinement.
}

A limitation of this study is the transverse system size.
A larger transverse-size along with the roughness would allow fluid domain to laterally bypass the asperity obstacles.
Thus, it will be interesting to investigate the level of persistence of plug-flow, cavitation, and folding-lip mechanisms when a lateral bypass pathway is available.

%
%
%
%
%



\section*{AUTHOR DECLARATIONS}

\subsection*{Conflict of Interest}
The authors have no conflicts to disclose.

\subsection*{Funding Statement}
This research did not receive funding.

\subsection*{Data Availability}
The LAMMPS input files, initial configurations including roughness realizations, interaction parameters, post-processing scripts, and data used to generate the principal figures are available at \url{https://github.com/shubham104084/Rough_lubrication_MD.git}. Additional trajectory files are available from the authors upon reasonable request because of their large size.

\subsection*{Acknowledgments}
Authors gratefully acknowledge Marc Honecker for helping with the initial set-up of the rough walls and 
Dr. Sergey Sukhomlinov for helpful discussions on selected LAMMPS related queries.

\appendix

\section{Material for Supplementary Material Section}

\subsection{Rotated elastic tensor}
\label{sec:stress_tensor}

Starting from the cubic stiffness tensor

\[
\mathbf C=
\begin{pmatrix}
C_{11}&C_{12}&C_{12}&0&0&0\\
C_{12}&C_{11}&C_{12}&0&0&0\\
C_{12}&C_{12}&C_{11}&0&0&0\\
0&0&0&C_{44}&0&0\\
0&0&0&0&C_{44}&0\\
0&0&0&0&0&C_{44}
\end{pmatrix},
\]
the tensor transformed to the
\(([1\bar{1}0],[11\bar{2}],[111])\) frame has trigonal (\(3m\)) form

\[
\mathbf C'=
\begin{pmatrix}
C'_{11} & C'_{12} & C'_{13} & C'_{14} & 0 & 0\\
C'_{12} & C'_{11} & C'_{13} & -C'_{14} & 0 & 0\\
C'_{13} & C'_{13} & C'_{33} & 0 & 0 & 0\\
C'_{14} & -C'_{14} & 0 & C'_{44} & 0 & 0\\
0 & 0 & 0 & 0 & C'_{44} & C'_{14}\\
0 & 0 & 0 & 0 & C'_{14} & C'_{66}
\end{pmatrix},
\]
with
$C'_{11}=(C_{11}+C_{12}+2C_{44})/2$, $C'_{12}=(C_{11}+5C_{12}-2C_{44})/6$, $C'_{13}=(C_{11}+2C_{12}-2C_{44})/3$, $C'_{14}=(C_{11}-C_{12}-2C_{44})/3\sqrt{2}$, $C'_{33}=(C_{11}+2C_{12}+4C_{44})/3$, $C'_{44}=(C_{11}-C_{12}+C_{44})/{3}$, and $C'_{66}=(C'_{11}-C'_{12})/2$. This is a standard result.


%







A $90^\circ$ in plane $x\leftrightarrow y$ rotation not only changes the sign of $C'_{14}$ but also transforms the stress-strain coupling coefficients. For the indenter (unrotated) the stress strain relation can we written as:
\[
\begin{pmatrix}
    \sigma_{x}\\\sigma_{y}\\\sigma_{z}\\\sigma_{yz}\\\sigma_{xz}\\\sigma_{xy}
\end{pmatrix}^{(A)} =
\begin{pmatrix}
C'_{11} & C'_{12} & C'_{13} & C'_{14} & 0 & 0\\
C'_{12} & C'_{11} & C'_{13} & -C'_{14} & 0 & 0\\
C'_{13} & C'_{13} & C'_{33} & 0 & 0 & 0\\
C'_{14} & -C'_{14} & 0 & C'_{44} & 0 & 0\\
0 & 0 & 0 & 0 & C'_{44} & C'_{14}\\
0 & 0 & 0 & 0 & C'_{14} & C'_{66}
\end{pmatrix}
\begin{pmatrix}
    \epsilon_{\text{inc}}\\-\epsilon_{\text{inc}}\\\epsilon_{z}\\2\epsilon_{yz}\\2\epsilon_{xz}\\2\epsilon_{xy}
\end{pmatrix},
\]
while that for rotated substrate can be written as
\[
\begin{pmatrix}
    \sigma_{x}\\\sigma_{y}\\\sigma_{z}\\\sigma_{yz}\\\sigma_{xz}\\\sigma_{xy}
\end{pmatrix}^{(B)} =
\begin{pmatrix}
C'_{11} & C'_{12} & C'_{13} & 0 & -C'_{14} & 0\\
C'_{12} & C'_{11} & C'_{13} & 0 & C'_{14} & 0\\
C'_{13} & C'_{13} & C'_{33} & 0 & 0 & 0\\
0 & 0 & 0 & C'_{44} & 0 & C'_{14}\\
-C'_{14} & C'_{14} & 0 & 0 & C'_{44} & 0\\
0 & 0 & 0 & C'_{14} & 0 & C'_{66}
\end{pmatrix}
\begin{pmatrix}
    -\epsilon_{\text{inc}}\\\epsilon_{\text{inc}}\\\epsilon_{z}\\2\epsilon_{yz}\\2\epsilon_{xz}\\2\epsilon_{xy}
\end{pmatrix},
\]
where $\epsilon_{\text{inc}}=\ln(s)$.
Therefore for indenter:
\[
\begin{pmatrix}
    \sigma_{x}-\sigma_{y}\\\sigma_{x}-\sigma_{z}\\\sigma_{y}-\sigma_{z}\\\sigma_{yz}\\\sigma_{xz}\\\sigma_{xy}
\end{pmatrix}^{(A)} =
\begin{pmatrix}
2(C'_{11}-C'_{12})\epsilon_{\text{inc}}  + 4C'_{14}\epsilon_{yz}\\
(C'_{11}-C'_{12})\epsilon_{\text{inc}}+(C'_{13}-C'_{33})\epsilon_{z}+2C'_{14}\epsilon_{yz}\\
-(C'_{11}-C'_{12})\epsilon_{\text{inc}}+(C'_{13}-C'_{33})\epsilon_{z}-2C'_{14}\epsilon_{yz}\\
2C'_{14}\epsilon_{\text{inc}}+2C'_{44}\epsilon_{yz} \\
2C'_{44}\epsilon_{xz}+2C'_{14}\epsilon_{xy}\\
2C'_{14}\epsilon_{xz}+2C'_{66}\epsilon_{xy}
\end{pmatrix},
\]
and for substrate
\[
\begin{pmatrix}
    \sigma_{x}-\sigma_{y}\\\sigma_{x}-\sigma_{z}\\\sigma_{y}-\sigma_{z}\\\sigma_{yz}\\\sigma_{xz}\\\sigma_{xy}
\end{pmatrix}^{(B)} =
\begin{pmatrix}
-(2(C'_{11}-C'_{12})\epsilon_{\text{inc}}  - 4C'_{14}\epsilon_{xz})\\
-(C'_{11}-C'_{12})\epsilon_{\text{inc}}+(C'_{13}-C'_{33})\epsilon_{z}-2C'_{14}\epsilon_{xz}\\
(C'_{11}-C'_{12})\epsilon_{\text{inc}}+(C'_{13}-C'_{33})\epsilon_{z}+2C'_{14}\epsilon_{xz}\\
2C'_{44}\epsilon_{yz}+2C'_{14}\epsilon_{xy} \\
2C'_{14}\epsilon_{\text{inc}}+2C'_{44}\epsilon_{xz}\\
2C'_{14}\epsilon_{yz}+2C'_{66}\epsilon_{xy}
\end{pmatrix},
\]
Therefore, even in the (unrealistically) simplified case of $\epsilon_{yz}^{(A)}\approx \epsilon_{xz}^{(B)}$, $\sigma_{\text{vM}}^{(A)}\neq\sigma_{\text{vM}}^{(B)}$, since $\epsilon_z\neq0$ in the loaded case.

\section{Sensitivity to roughness realization}
\label{sec:rough_real_2}

\hl{To examine whether the principal lubrication mechanisms identified in the main simulations depend on the particular asperity configuration, we performed an additional set of simulations at the sliding-speed of 50~m/s for the three lubricants using an independently generated roughness realization with same prescribe roughness statistics.}

\begin{figure*}[h!]
    \centering
    \includegraphics[width=0.85\textwidth]{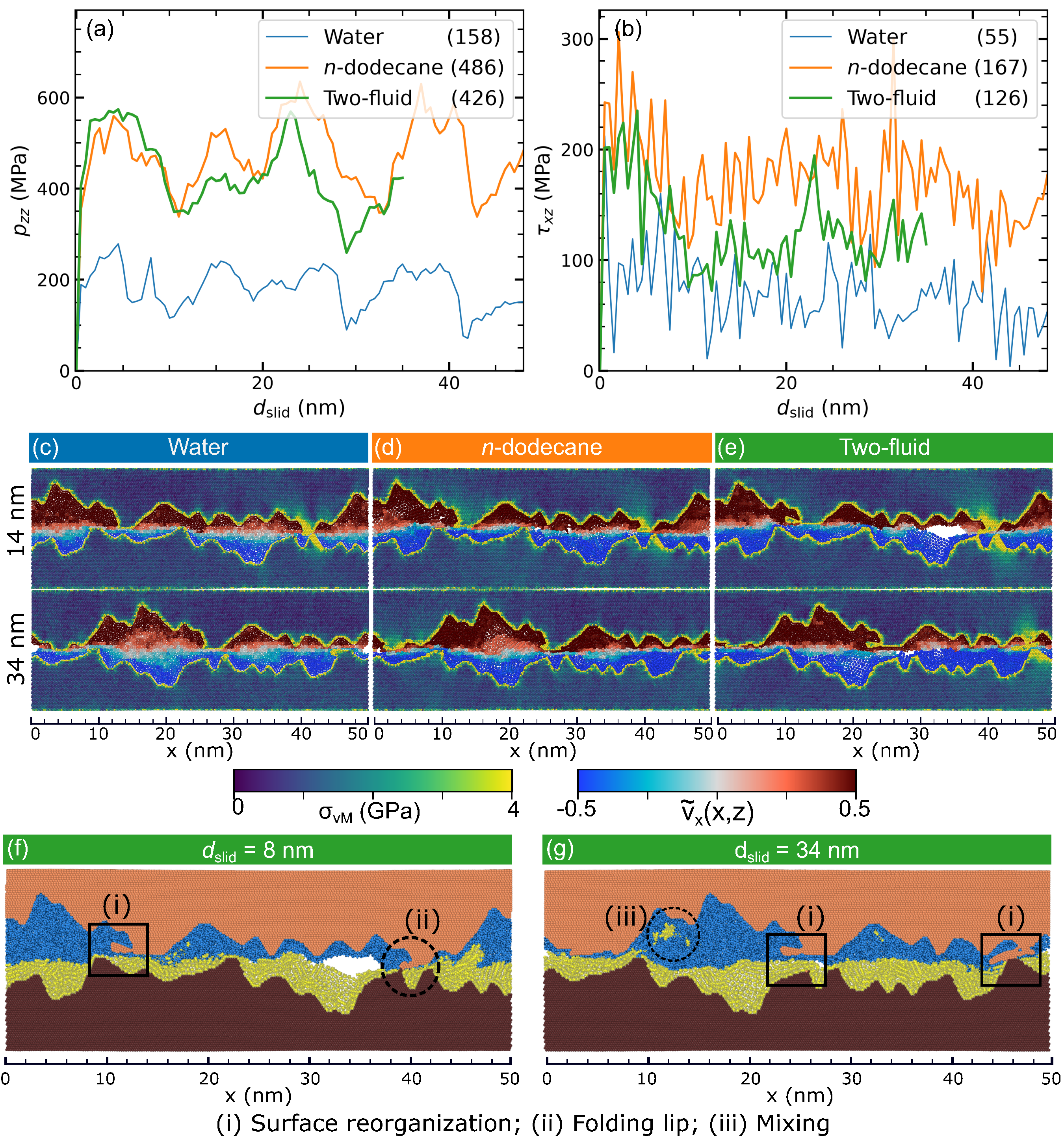}
    \caption{Response of the three lubricant systems at \(v=50\) m/s for an independent roughness realization. 
    Panels (a)--(e) follow the set-up in Fig.~\ref{fig:water_vel}.  
    %
    Panels (f) and (g) show representative two-fluid configurations highlighting (i) surface reorganization, (ii) folding-lip formation, and (iii) transient local mixing following asperity engagement.
    \label{fig:RR2}
    }
\end{figure*}

\hl{\sa{Fig.~\ref{fig:RR2}} compares the resulting normal and shear stresses. 
Although the detailed stress histories and mean stress values differ from those obtained for the primary roughness realization, the principal lubricant-dependent behavior is retained.
Water again supports substantially less normal load than \(n\)-dodecane, whereas the normal-load support of the two-fluid system remains much closer to that of \(n\)-dodecane than to water.
Such quantitative differences are expected, because changing the random realization alters the local gap distribution and therefore the effective confinement of the lubricant, even when the statistical roughness parameters are unchanged. 
The purpose of this additional realization is therefore not to quantify error bars, but to examine whether the principal lubricant-dependent mechanisms persist for a different asperity configuration.}

\hl{
Representative configurations are shown in \sa{Fig.~\ref{fig:RR2}(c--e)}.
Water again undergoes substantial squeeze-out and cavitation during roughness-driven asperity interactions, whereas \(n\)-dodecane retains a more persistent confined film. The two-fluid system shows an intermediate and dynamically evolving response. In addition to maintaining substantial fluid-supported load, it again exhibits pronounced interfacial and surface reorganization.
The latter behavior is illustrated more clearly in \sa{Fig.~\ref{fig:RR2}(f,g)}. Asperity engagement locally reorganizes the solid–fluid interface and produces folding-lip structures similar to those observed for the primary realization. Transient local mixing of water and \(n\)-dodecane is also observed following strong asperity engagement, after which the two phases largely recover their preferential association with the opposing surfaces. The recurrence of both folding-lip formation and transient mixing indicates that these features are not specific to the particular asperity arrangement of the primary realization.
}

\hl{The second realization additionally provides a possible explanation for the condition-dependent frictional response of the two-fluid lubricant.
A few events of strong asperity engagement and the accompanying surface reorganization alter the subsequent local topography and produce regions of increased separation.
This partially relieves the confinement of the lubricant, which weakens the strongly localized shear observed in the single-component lubricants, whereas appreciable relative motion remains concentrated near the \(n\)-dodecane–water interface in the two-fluid case.
This difference in shear accommodation may explain the lower friction of the two-fluid lubricant in the second realization.}

\hl{
The second realization supports the qualitative robustness of the principal mechanisms identified in the main simulations, while also showing that their quantitative manifestation and the resulting frictional response can depend on the roughness-induced local confinement.}

\section{Effect of wall geometry on directional material transfer}
\label{sec:app_swap_wall}

\begin{figure*}[hbt!]
    \centering
    \includegraphics[width=0.95\textwidth]{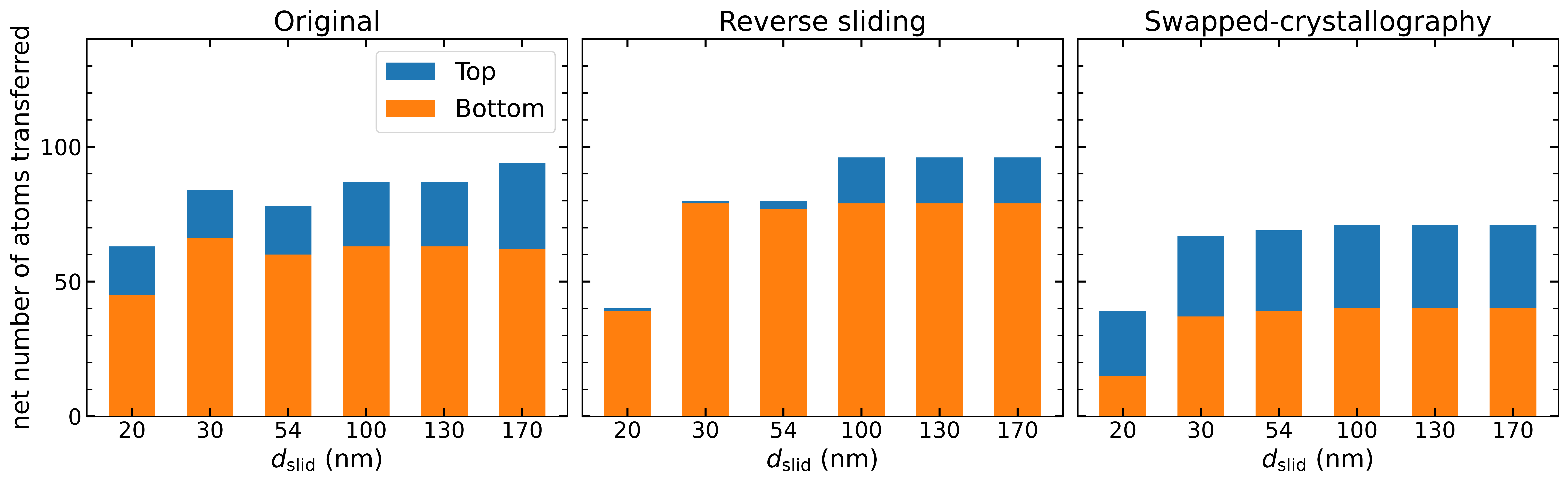}
    \caption{Net material transfer at 50 m/s with $n$-dodecane for the original, reverse-sliding, and swapped crystallographic configurations. “Top” denotes atoms transferred from the top to the bottom wall, and “Bottom” vice versa. Swapping the crystallographic configurations substantially reduces the directional bias without reversing it.
    \label{fig:wear_swap_wall}
    }
\end{figure*}

\hl{To examine whether the directional material transfer is influenced by the asymmetric crystallographic configuration of the two copper blocks, we performed an additional control simulation at 50~m/s for $n$-dodecane in which the crystallographic configurations of the two walls were swapped while keeping the remaining geometry and simulation conditions unchanged.
Figure \ref{fig:wear_swap_wall} compares the original, reverse-sliding, and swapped configurations.
In the original configuration, material transfer is strongly biased from bottom surface to top surface.
Sliding in the reverse direction retains this bias.
However, after swapping the wall configurations, this directional bias is substantially reduced but not reversed.
For instance, at a sliding distance of 100~nm, the excess transfer decreases from 39 to 9 atoms, corresponding to a 77~\% reduction in directional bias.
This shows that the crystallographic asymmetry contributes substantially to the preferred transfer direction, but does not determine it alone. The remaining asymmetry may also reflect the geometric inequivalence of the two surfaces, since deterministic curvature is imposed only on the top surface.}

\hl{Reversing the sliding velocity in the original system changes the values but not the trend. 
Now atoms initially transfer almost exclusively from the bottom, which is soft in the sliding direction, to the top wall, which is stiff in the lateral direction.
However, after sliding 170~nm, the disparity between the two values diminishes.
}

\bibliographystyle{elsarticle-num}
\bibliography{bib}

\end{document}